%% file: paper.tex
\pdfoutput=1
\documentclass[nonacm,acmlarge,authorversion]{acmart} 

\usepackage[utf8]{inputenc} % allow utf-8 input
\usepackage[T1]{fontenc}    % use 8-bit T1 fonts
\usepackage{url}
\usepackage{latexsym}
\usepackage{enumitem}
\usepackage{color, colortbl}
\usepackage{threeparttable,booktabs}

\usepackage[hang,flushmargin]{footmisc}

\usepackage{booktabs}
\usepackage{multirow}
\usepackage{adjustbox}
\usepackage{arydshln}
\usepackage{footnote}
\usepackage{tabularx}
\usepackage{array}
\usepackage{graphicx,subcaption}
\usepackage{enumitem}
\usepackage{amsmath}
\usepackage{amsfonts}
\usepackage{dsfont}
\usepackage{pifont}% http://ctan.org/pkg/pifont

\newcommand{\cmark}{\ding{51}}%
\newcommand{\xmark}{\ding{55}}%

\newcommand{\qbow}{\mathbf{q}_{\textrm{\tiny BoW}}}
\newcommand{\dbow}{\mathbf{d}_{\textrm{\tiny BoW}}}

\newcommand{\qdlr}{\mathbf{q}_{\textrm{\tiny DLR}}}
\newcommand{\ddlr}{\mathbf{d}_{\textrm{\tiny DLR}}}
\newcommand{\gdlr}{\mathbf{g}_{\textrm{\tiny DLR}}}
\newcommand{\qdhr}{\mathbf{q}_{\textrm{\tiny DHR}}}
\newcommand{\ddhr}{\mathbf{d}_{\textrm{\tiny DHR}}}
\newcommand{\gdhr}{\mathbf{g}_{\textrm{\tiny DHR}}}

\newcommand{\qcls}{\mathbf{q}_{\texttt{{\tiny [CLS]}}}}
\newcommand{\dcls}{\mathbf{d}_{\texttt{{\tiny [CLS]}}}}
\newcommand{\spladedlr}{\text{SPLADE}_{\text{\tiny DLR}}}
\newcommand{\deladedlr}{\text{DeLADE}_{\text{\tiny DLR}}}
\newcommand{\unicoildlr}{\text{uniCOIL}_{\text{\tiny DLR}}}
\newcommand{\deladedhr}{(\text{DeLADE+[CLS]})_{\text{\tiny DHR}}}

\definecolor{bluencs}{rgb}{0.0, 0.53, 0.74}

\definecolor{Gray}{gray}{0.9}
\definecolor{pink}{rgb}{0.858, 0.188, 0.478}
\definecolor{bleudefrance}{rgb}{0.19, 0.55, 0.91}

\begin{document}
\title{A Dense Representation Framework for Lexical and Semantic Matching}

\author{Sheng-Chieh Lin}
\affiliation{%
%   \institution{\vskip .1cm}%add some spacing if needed
  \institution{University of Waterloo}
  \city{Waterloo}
 \country{Canada}
}
\author{Jimmy Lin}
\affiliation{%
%   \institution{\vskip .1cm}%add some spacing if needed
  \institution{University of Waterloo}
   \city{Waterloo}
 \country{Canada}
}

\renewcommand{\authors}{Sheng-Chieh Lin, Jimmy Lin}

\begin{abstract}
Lexical and semantic matching capture different successful approaches to text retrieval and the fusion of their results has proven to be more effective and robust than either alone. 
Prior work performs hybrid retrieval by conducting lexical and semantic matching using different systems (e.g., Lucene and Faiss, respectively) and then fusing their model outputs. 
In contrast, our work integrates lexical representations with dense semantic representations by densifying high-dimensional lexical representations into what we call low-dimensional dense lexical representations (DLRs). 
Our experiments show that DLRs can effectively approximate the original lexical representations, preserving effectiveness while improving query latency. 
Furthermore, we can combine dense lexical and semantic representations to generate dense hybrid representations (DHRs) that are more flexible and yield faster retrieval compared to existing hybrid techniques.
In addition, we explore {\it jointly} training lexical and semantic representations in a single model and empirically show that the resulting DHRs are able to combine the advantages of the individual components. 
Our best DHR model is competitive with state-of-the-art single-vector and multi-vector dense retrievers in both in-domain and zero-shot evaluation settings.
Furthermore, our model is both faster and requires smaller indexes, making our dense representation framework an attractive approach to text retrieval. 
Our code is available at \url{https://github.com/castorini/dhr}.
\end{abstract}




\maketitle

\section{Introduction}
\label{sec:intro}

Transformer-based bi-encoders have been widely used as first-stage retrievers for text retrieval. 
Compared to their multi-vector counterparts~\cite{colbert, coil, colberter}, single-vector representation learning approaches (with a few representative techniques listed in Table~\ref{tb:summary}) are attractive due to their good balance between effectiveness and efficiency. 

Semantic matching through dense representations~\cite{sentence-bert, Chang2020Pre-training, dpr} form one large successful class of models.\footnote{In the literature, these are often just called dense retrieval models. However, we explicitly refer to these as dense {\it semantic} representations because we show that dense {\it lexical} representations exist as a separate class of models.} 
These dense semantic representations transform the relevance matching problem into nearest neighbor search in a semantic space, tackling vocabulary and semantic mismatches in ways that traditional lexical matching approaches (e.g., BM25) cannot.
Subsequent work further improves these models through advanced training techniques~\cite{tct, tasb, xiong2020approximate, star, condenser, contriever, rocketqa}. 
However, as \citet{sciavolino2021simple} show, dense semantic representations still fail in some easy cases, and it remains challenging to interpret why they sometimes perform poorly. 
In addition, \citet{beir} demonstrate that many existing dense retrievers still fall short in terms of generalization capability across different domains. 

There is evidence~\cite{mebert, clear, unicoil} that lexical matching compensates for the weaknesses of semantic matching; these papers further propose hybrid retrieval techniques that fuse lexical and semantic representations. 
However, in practice, lexical and semantic matching are executed in very different ways:\ typically, lexical matching is conducted using inverted indexes, for example, in Lucene, while semantic matching is treated as nearest neighbor search using, for example, HNSW indexes in Faiss~\cite{faiss}.
This means that a hybrid retrieval system requires two separate ``software stacks'', running completely distinct retrieval operations in parallel before their outputs are post-processed to generate a final ranking (e.g., through linear combination of scores).
Such a design makes real-world deployments more complicated than necessary, since, for example, the two separate indexes need to be maintained and be kept in sync.

Recently, another thread of work uses bi-encoders to learn sparse lexical (bag-of-words) representations for text retrieval. 
For example, \citet{deepct} demonstrate with DeepCT that replacing tf--idf with contextualized term weights from a transformer-based regression model significantly improves retrieval effectiveness. 
Subsequent work further combines term reweighting techniques with term expansion to address vocabulary and semantic mismatch issues with lexical representations.
Some methods~\cite{deepimpact, unicoil, tildev2} leverage another model for expansion, which incurs additional costs in both training and inference. 
As an alternative, \citet{splade-v2} exploit BERT's masked language model (MLM) layer to train a single model for both expansion and term weighting. 

Compared to dense semantic models, sparse lexical models appear to be more robust to domain shifts~\cite{beir}.
However, the optimization of sparse lexical retrievers must take into account the efficiency of query evaluation using inverted indexes. 
For example, to achieve better effectiveness, term expansion techniques tend to make sparse lexical representations more dense, sometimes rendering retrieval with inverted indexes much slower~\cite{Mackenzie_etal_arXiv2021}. 
Further performance penalties are incurred when integrating sparse lexical and dense semantic representations into hybrid retrieval systems. 
Since it is impractical to directly compute dot products between high-dimensional vectors at scale in latency-sensitive retrieval applications (i.e., via brute-force approaches), using lexical representations with inverted indexes remains presently the only sensible choice, but this comes with the aforementioned limitations.  

\begin{figure}[t]
    \centering
    \resizebox{0.75\columnwidth}{!}{
        \includegraphics{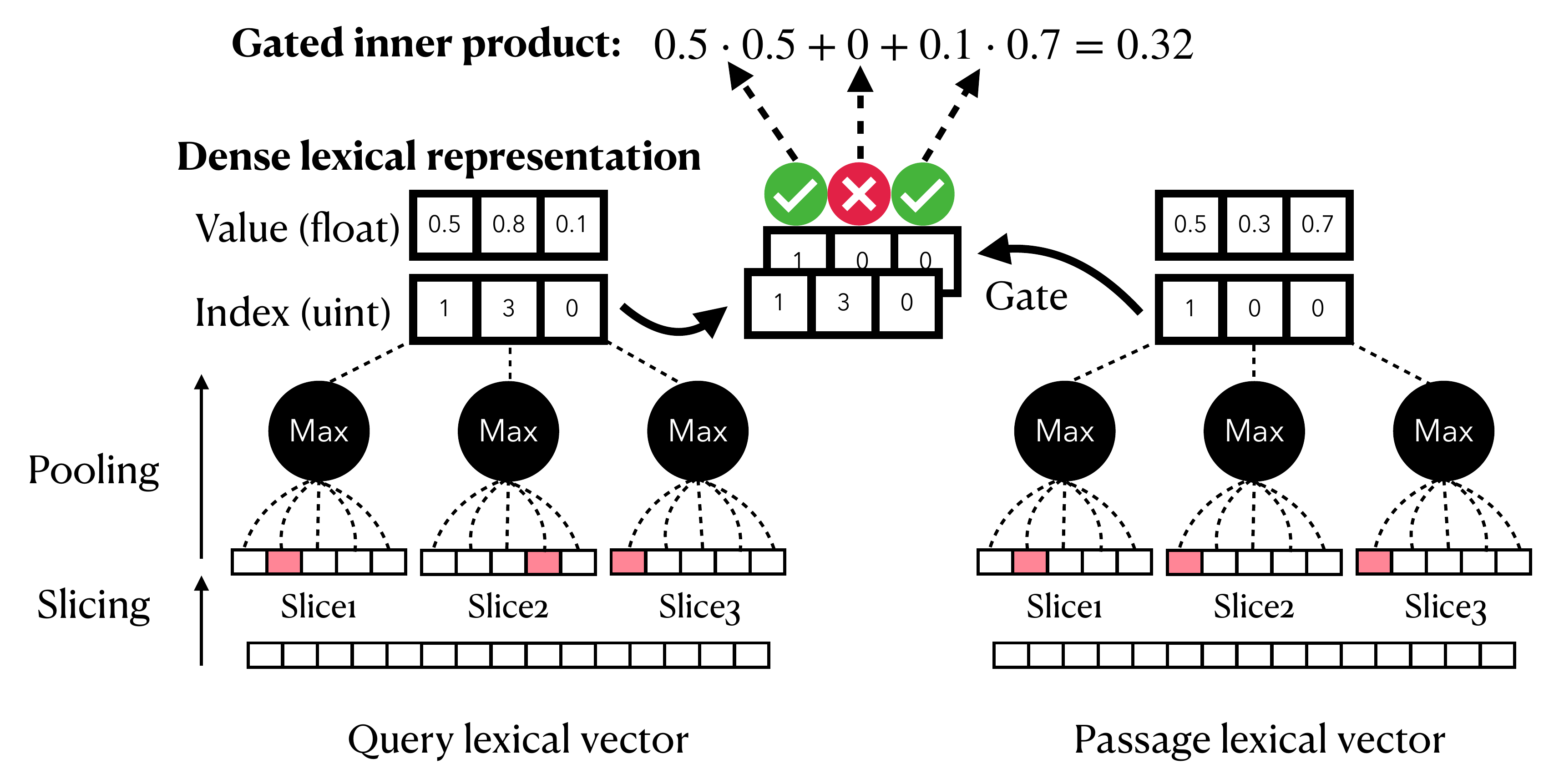}
    }
    \caption{Illustration of densified lexical representations (DLRs) and gated inner product (GIP). Both query and passage lexical representations are fixed-width vectors where the number of dimensions is equal to the vocabulary size. Our approach first groups these high-dimensional vectors into $M$ slices, each with $N$ dimensions (e.g., $M=3$, $N=5$ here). For each slice, the maximum value is selected. These values from the original vector (Value) and their positions in each slice (Index) are recorded separately. When computing the gated inner product between two vectors, only the dimensions with the same index are considered.}
    \label{fig:idea}
\end{figure}

Motivated by these tradeoffs, we explore an approach to {\it directly} computing dot products between lexical vectors in a scalable and low-latency manner for retrieval applications. 
This is accomplished by densifying lexical representations, which can be applied to any existing lexical model.
The approach described in this paper, which builds on our previous work~\cite{dhrv1}, is comprised of two simple steps:\ (1) representation slicing (2) sliced representation pooling. 
This method can be viewed as compressing high-dimensional lexical representations into low-dimensional ones, as depicted in Fig.~\ref{fig:idea}. 
One key feature of our approach is that it does not involve either unsupervised~\cite{faiss} or supervised training~\cite{jpq, RepCONC, spar}.   
Any collection of lexical representations (from a ``base model'') can be converted into dense lexical representations (DLRs), and the nearest neighbor search problem between query and passage vectors can be performed with an operation we call the gated inner product (GIP). 
Standard dense representations and inner products can be considered a special case of DLRs and GIP, respectively. 
Thus, our representation and scoring function comprise a unified framework for {\it both} lexical and semantic matching.

An advantage of our framework is that end-to-end lexical and hybrid retrieval can be performed {\it directly} on GPUs with a single index structure in a unified execution environment.
In fact, our implementation uses standard vector operations directly in the popular PyTorch open-source neural modeling toolkit.
Retrieval efficiency, unlike with inverted indexes, is not sensitive to the sparsity of representations; thus, we can optimize retrieval effectiveness without any constraints. 
Building on these features, we propose a new model called Dense Lexical AnD Expansion (DeLADE) and explore a new matching approach, which we refer to as dense lexical matching (shown in Table~\ref{tb:summary}). 

In addition, we can integrate DLRs with dense semantic representations into what we call dense hybrid representations (DHRs).
This can be accomplished in two ways:\ First, we can independently combine DLRs with any ``off-the-shelf'' dense semantic representations such as ANCE~\cite{xiong2020approximate}, TAS-B~\cite{tasb}, etc.
Second, we can {\it jointly} train DHRs that combine lexical as well as semantic components. 
Both DLRs and DHRs exhibit advantageous features for real-world applications:\ (1) end-to-end retrieval can be accomplished on GPUs using a single index and software framework, instead of, for example, using Lucene for lexical matching and Faiss for semantic matching; (2) vector densification can be applied to meet storage constraints without model retraining. 

With our proposed dense representation framework, we explore the following research questions:

\newcommand{\RQone}{\begin{itemize}
    \item[\textbf{RQ1}] How well do DLRs approximate the original high-dimensional lexical representations? 
\end{itemize}}
\newcommand{\RQoneRunning}{\textbf{RQ1} \textit{..? }}
\RQone

\noindent We densify lexical representations from two lexical models based on whole word matching (BM25 and DeepImpact~\cite{deepimpact}) and two lexical models based on wordpiece token matching (uniCOIL~\cite{unicoil} and SPLADE~\cite{splade-v2}). 
Experimental results show that our approach effectively compresses high-dimensional lexical representations (30K and even 3.5M dimensions) into 768- and 128-dimensional vectors with less than 1$\%$ and 5$\%$ retrieval effectiveness loss, respectively. 
In addition, this approach can be applied to DeLADE, our proposed dense lexical representation model, which is a SPLADE variant. 
More importantly, this compression enables us to perform lexical matching directly on GPUs, which substantially speeds up retrieval compared to using an inverted index and does not depend on the sparsity of the lexical representations. 

\newcommand{\RQtwo}{\begin{itemize}
    \item[\textbf{RQ2}] How well do DHRs benefit from the independent fusion of DLRs and ``off-the-shelf'' dense semantic representations?
\end{itemize}}
\newcommand{\RQtwoRunning}{\textbf{RQ2} \textit{..? }}
\RQtwo

\noindent Within our framework, we propose hybrid dense representations (DHRs) by combining DLRs and other standard ``off-the-shelf'' dense semantic representations for hybrid fusion retrieval. 
We demonstrate that our method achieves comparable retrieval effectiveness to other existing hybrid retrieval methods but with lower query latency.  

Next, we propose to {\it jointly} train lexical and semantic components within DHRs. 
Specifically, we combine lexical representations and the [CLS] embeddings from a transformer to capture lexical and semantic matching, and then conduct retrieval based on a fusion of these separate components in a single unified framework.

\newcommand{\RQthree}{\begin{itemize}
    \item[\textbf{RQ3}] Can DHRs benefit from {\it joint} training of lexical and semantic components in a single model?
\end{itemize}}
\newcommand{\RQthreeRunning}{\textbf{RQ3} \textit{..? }}
\RQthree

\noindent Experiments show that our approach to jointly training DHRs (i.e., single model fusion) outperforms both dense semantic and sparse lexical retrieval models and is competitive with state-of-the-art multi-vector approaches such as ColBERT~\cite{colbert} and COIL~\cite{coil}.
We achieve effectiveness on par with the state of the art on an in-domain benchmark (MS MARCO~\cite{marco}) and obtain better generalization on an out-of-domain benchmark (BEIR~\cite{beir}). 
In addition, the index size of our model can be tuned by controlling the vector dimensionality of the lexical component {\it without retraining}, making our approach attractive for real-world applications. 

Further analysis identifies that GIP requires more operations compared to the standard inner product for dense vectors and this is one potential drawback of end-to-end retrieval with DLRs. 
To address this issue, we propose a two-stage retrieval approach:\ in the first stage, retrieval is conducted by {\it approximate} GIP, which is faster but less accurate; then, the retrieved sets are reranked by computing the exact GIP. 
Thus, we explore the following research question:

\newcommand{\RQfour}{\begin{itemize}
    \item[\textbf{RQ4}] How effective is our proposed two-stage retrieval approach?
\end{itemize}}
\newcommand{\RQfourRunning}{\textbf{RQ4} \textit{..? }}
\RQfour

\noindent Our experiments show that approximate GIP is capable of retrieving sufficient relevant passages (i.e., achieves good recall) for the subsequent exact GIP reranking to achieve high end-to-end effectiveness. 
Specifically, we demonstrate that with approximate GIP, our proposed two-stage retrieval approach substantially reduces retrieval latency without sacrificing any effectiveness compared to end-to-end retrieval using more expensive but exact GIP. 

Our contributions are summarized as follows:

\begin{itemize}

\item We propose a simple yet effective approach to densifying high-dimensional lexical representations for text retrieval, creating what we call dense lexical representations (DLRs).

\item Building on DLRs, we introduce dense hybrid representations (DHRs) that combine lexical and semantic representations.

\item While DHRs can combine arbitrary off-the-shelf lexical representations (e.g., BM25 and uniCOIL) and semantic representations (e.g., ANCE and TAS-B) {\it independently}, we demonstrate how to {\it jointly} train effective DHRs with complementary lexical and semantic components.

\item We show how to efficiently conduct two-stage retrieval in our dense representation framework, with fast approximate GIP followed by exact GIP reranking.

\end{itemize}

\noindent Code to reproduce all experiments in this paper is available at \url{https://github.com/castorini/dhr}. 

\begin{table}[t]
\caption{Comparison of Single-Vector Representation Learning Approaches}
\label{tb:summary}
	\centering
	\begin{small}
    \begin{tabular}{ll!{\color{black}\vrule}c!{\color{black}\vrule}c}
	\toprule
 \multicolumn{2}{c}{matching type}& \multicolumn{1}{c}{Sparse}  &  Dense   \\
    \midrule
    \multirow{2}{*}{Lexical} & whole words& DeepCT~\cite{deepct}, DeepImpact~\cite{deepimpact}  & \multirow{2}{*}{ \textbf{DLR} (our approach)} 
    \\
    \cline{2-3}
    & wordpiece tokens & uniCOIL~\cite{unicoil}, SPLADE~\cite{splade} \\
     \midrule
    \multirow{2}{*}{Semantic} & & \multirow{2}{*}{UHD~\cite{ultra}}& DPR~\cite{dpr}, ANCE~\cite{xiong2020approximate}\\
    & & & TASB~\cite{tasb}, RocketQA~\cite{rocketqa} \\
    \midrule
    Hybrid && -& \textbf{DHR} (our approach)\\
         \bottomrule
	\end{tabular}
	\end{small}
\end{table}

\section{Background and Related Work}
\label{sec:backgroud}

Following \citet{lin2020pretrained}, let us formulate the task of text (or \textit{ad hoc}) retrieval as follows:\ Given a query $q$, the goal is to retrieve a ranked list of documents $\{d_1, d_2, \cdots d_k\} \in C$ to maximize some ranking metric, such as nDCG, where $C$ is the collection of documents.  

Specifically, given a (query, passage) pair, we aim to maximize the following:
\begin{align}
    \textrm{sim}(q,d) \triangleq \phi(\eta_q(q), \eta_d(d)) =  \langle \mathbf{q}, \mathbf{d} \rangle, 
\end{align}
\noindent where $\eta_q(\cdot)$ and $\eta_d(\cdot) \in \mathbb{R}^h$ denote functions mapping the query and the passage into $h$-dimensional vector representations, $\mathbf{q}$ and $\mathbf{d}$, respectively.
The scoring function that quantifies the degree of relevance between the representations $\mathbf{q}$ and $\mathbf{d}$ is denoted $\phi(\cdot,\cdot)$, which can be a simple inner product or a more complex operation~\cite{colbert, coil, marco_BERT, poly-encoders}. 
We focus on single-vector representation learning approaches that apply the inner product as the scoring function. 
We categorize single-vector representation learning approaches through ``matching type'', as shown in Table~\ref{tb:summary}. 
In the literature, there are two main lines of research:\ dense representations for semantic matching and sparse representations for lexical matching. 

\paragraph{Dense representations for semantic matching.}
Pretrained transformers~\cite{devlin2018bert, liu2019roberta} are able to encode sentences or passages into dense semantic representations, which have been shown to be effective for downstream tasks~\cite{sentence-bert, Chang2020Pre-training}. 
In recent years, transformer-based bi-encoders have been widely applied to the task of passage retrieval~\cite{dpr} and further improved by advanced training techniques such as hard negative mining~\cite{xiong2020approximate,star}, knowledge distillation~\cite{tasb, tct}, pretraining~\cite{condenser, contriever}, or their combination~\cite{rocketqa}.
These approaches encode queries and passages into dense vectors, using the inner product to capture the degree of relevance:
\begin{align}
     \textrm{sim}_{\textrm{\tiny semantic}}(q,d) \triangleq  \langle \qcls, \dcls \rangle, 
\end{align}
\noindent where $\qcls$ and $\dcls$ are typically 768-dimensional vectors taken from the [CLS] token in the final layer of a transformer model (or alternatively, pooling over the contextualized representations of the tokens).

\paragraph{Sparse representations for lexical matching.} 
To our knowledge, \citet{SNRM} was the first to demonstrate that neural networks can learn lexical representations for text retrieval.
Recently, transformer-based bi-encoders have also been applied to lexical representation learning by replacing heuristic term weighting functions (e.g., BM25) with learned term weights.
In the literature, these solutions can be classified into two broad classes:\ (1) linear layer and (2) MLM projection.

One early technique, DeepCT~\cite{deepct} uses a linear layer to project BERT token embeddings into contextualized term weights for input tokens.
Subsequent work~\cite{deepimpact,unicoil,tildev2} further combines DeepCT and other document expansion methods~\cite{doctttttquery,tilde} to address vocabulary mismatch and missing terms in DeepCT. 
For example, DeepImpact~\cite{deepimpact} uses a trained sequence-to-sequence model~\cite{doctttttquery} to expand the original passages in the corpus and learn contextualized term weights. 
Other models~\cite{unicoil,tildev2} follow a similar approach to DeepImpact but term matching is performed in the space of BERT wordpiece tokens rather than whole words. 
They generate term weights only for tokens appearing in each (possibly expanded) query or passage and thus the lexical representations are sparse by design.
However, to achieve competitive effectiveness, these techniques require additional models for term expansion (see \citet{tildev2} for more discussion).
In contrast, some researchers~\cite{splade, sparterm} use the masked language model (MLM) layer in transformers such as BERT to learn term weighting and expansions at the same time.
For these approaches, sparsity regularization must be applied during model training to ensure that the generated lexical representations are amenable to retrieval using inverted indexes.

Generally, these neural approaches to lexical term matching can be viewed as projecting queries and passages into $|V_{\text{BERT}}|$-dimensional vectors, where $|V_{\text{BERT}}|=30522$ is the vocabulary size of BERT wordpiece tokens:
\begin{align}
    \textrm{sim}_{\textrm{\tiny lexical}}(q,d) \triangleq  \langle \qbow, \dbow \rangle, 
\end{align}
where $\mathbf{q}_{\textrm{\scriptsize BoW}}$ and  $\mathbf{d}_{\textrm{\scriptsize BoW}}\in \mathbb{R}^{30522}$.
The value in each dimension is the token's term weight.
As with dense retrieval, the relevance score between a query and a passage is computed by the inner product of their vector representations.

\paragraph{Bridging the gap between the two worlds.}
Although the inner product is a common operation for capturing relevance in the aforementioned two approaches, there are still two major differences between them:

\begin{enumerate}
\item Unlike semantic representations, lexical representations can be considered bags of words (or subwords) and thus are more interpretable, since dimensions of the representation vectors directly correspond to vocabulary items.

\item Text retrieval using lexical representations is usually performed using standard inverted indexes due to their high dimensionality, while text retrieval using semantic representations is usually performed using completely different infrastructure, e.g., HNSW indexes. 

\end{enumerate}

\noindent Previous work~\cite{mebert, clear, unicoil} has demonstrated that semantic and lexical matching can compensate for each other; these papers typically implement dense--sparse hybrid retrieval by performing retrieval independently using different systems and then merging their results (e.g., by interpolating scores). 
To make such an approach ``production-ready'' for deployment in real-world applications, non-trivial software engineering effort is required to coordinate dense and sparse retrieval in parallel (on inverted and HNSW indexes) and the final fusion. 
The need for two entirely separate ``software stacks'' and the associated operational maintenance costs (e.g., of keeping indexes in sync) increase the complexity of hybrid retrieval systems.

To bridge the gap between semantic and lexical representations for text retrieval, some researchers extend their focus beyond the above two research threads. 
For example, \citet{ultra} show that semantic matching can be executed using an inverted index by projecting semantic representations from BERT to sparse representations in an ultra-high-dimensional space. 
However, this projection operation requires additional training and the retrieval effectiveness of this approach still lags behind baseline dense retrieval approaches.

\begin{table}[t]
\caption{Comparison of Different Vector Compression Approaches}
\label{tb:compression_comparison}
	\centering
	\begin{small}
    \begin{tabular}{lcc}%!{\color{black}\vrule}cc}
	\toprule
 \multicolumn{1}{l}{Approach}& \multicolumn{1}{c}{unsupervised training}  &  \multicolumn{1}{c}{supervised training}   \\
    \midrule
    PQ~\cite{PQ}, OPQ~\cite{OPQ}, LSH~\cite{lsh}& \cmark& \xmark\\
    JPQ~\cite{jpq}, RepCONC~\cite{RepCONC}& \cmark& \cmark \\
    SPAR~\cite{spar}& \xmark& \cmark \\
    Our work& \xmark& \xmark \\
         \bottomrule
	\end{tabular}
	\end{small}

\vspace{0.25cm}

{\small The advantage of our approach is that it does not require {\it any} training.}
\end{table}

In contrast, another approach to bridging dense and sparse representations is to compress high-dimensional sparse lexical representations into low-dimensional dense ones. 
We can accomplish this using existing unsupervised approaches such as product quantization (PQ)~\cite{PQ}, optimized product quantization (OPQ)~\cite{OPQ}, and locality-sensitive hashing (LSH)~\cite{lsh}. 
However, such approaches require substantial computational resources since the entire corpus (or large portions thereof) need to be loaded into CPU/GPU memory to perform the unsupervised training. 
These approaches are not practical to compress ultra-high-dimensional lexical vectors, especially when the corpus is large.\footnote{As an example, BM25 with a standard English tokenizer in the Lucene search library generates representations with 2.6M dimensions for the MS MARCO passage corpus (containing 8.8M passages).}
Supervised compression techniques~\cite{jpq, RepCONC} that are built upon these unsupervised approaches similarly suffer from high resource requirements.

Closest to our own work, \citet{spar} distill lexical matching signals from existing sparse retrieval models such as BM25 and uniCOIL~\cite{unicoil} into low-dimensional dense representations in an approach called SPAR. 
Nevertheless, SPAR requires massive amounts of training data and computational resources (e.g., 64 V100 GPUs for three days). 
In contrast, our approach simply performs max pooling over each ultra-high-dimensional lexical vector.
Thus, we do not require any additional computational resources for unsupervised or supervised training. 
We summarize the comparison of different compression techniques in Table~\ref{tb:compression_comparison}.

Beyond not needing any training (either supervised or unsupervised), our approach to lexical representation compression confers the additional advantage of interpretability.
Specifically, we densify lexical representations in a reversible manner (except for some information loss from the max pooling operation). 
Thus, our dense lexical vectors still retain characteristics of the original lexical representations with respect to matching terms, as we will demonstrate.

\section{Methodology}
\label{sec:method}

In this section, we first describe our approach to densifying lexical representations by slicing and pooling. 
We then introduce our dense representation framework, under which the densified lexical representations are captured in compact pairs of vectors.
We then propose a new scoring function called gated inner product (GIP) for computing query--passage similarity. 
Next, we describe how to combine lexical and semantic representations using our framework and propose a two-stage approach for end-to-end retrieval. 
Finally, we introduce our dense lexical model, DeLADE.

\subsection{Dense Lexical Representations}

Lexical representations can be considered vectors with $|V|$ dimensions, i.e., $\qbow =(q_0,\cdots q_{|V|-1})$ and $\dbow =(d_0, \cdots d_{|V|-1})$.
These representations come from an underlying ``base model'' such as uniCOIL or SPLADE, and our focus here is to ``densify'' such representations.

We first divide each vector into $M$ slices, each of which is a smaller vector with $N$ dimensions (i.e., $|V|=M \cdot N$).
In terms of the standard ``slice'' notation used by Python:
\begin{align}
\label{eq:contiguous_slicing}
    &S^{q}_m = \qbow[mN:mN+N] \in \mathbb{R}^{N}; \nonumber \\ 
    &S^{d}_{m} = \dbow[mN:mN+N] \in \mathbb{R}^{N},
\end{align}
\noindent where $m \in \{0,1,\cdots,M-1 \}$. 
Note that the slicing can be performed in different ways; for example, slicing randomly or with a fixed stride: $[m:M(N-1)+m:N]$. 
For simplicity, we use contiguous slicing as shown in Eq.~(\ref{eq:contiguous_slicing}) in our presentation.
Thus, the inner product between $\qbow$ and $\dbow$ can be rewritten as the summation of all the dot products of their slices:
\begin{align}
\label{eq:bucket_dot}
    \langle \qbow, \dbow\rangle =  \sum_{m=0}^{M-1} \langle S^{q}_m , S^{d}_m \rangle.
\end{align}
\noindent Intuitively, if a lexical representation is sparse enough, we can assume that for each slice, there is only one non-zero entry. 
Thus, we can approximate $S^{q}_m$ ($S^{d}_m$) by keeping only the entry with the maximum value in each slice:    
\begin{align}
    S^{q}_m \approx \max{(S^{q}_m)} \cdot \hat{\mathbf{u}}(\text{e}^{q}_m); \nonumber \\
     S^{d}_m \approx \max{(S^{d}_m)} \cdot \hat{\mathbf{u}}(\text{e}^{d}_m),
\end{align}
where $\hat{\mathbf{u}}(\text{e}^q_m)$ is a unit vector with the only non-zero entry at the entry $\text{e}^{q}_{m}=\mathrm{argmax}{(S^{q}_{m})}$.
Thus, the inner product of $\qbow$ and $\dbow$ lexical vectors in Eq.~(\ref{eq:bucket_dot}) can be approximated as follows:
\begin{align}
\label{eq:approximate_dot}
    \langle\qbow, \dbow\rangle &\approx 
    \sum_{m=0}^{M-1}  \max{(S^{q}_m)} \cdot \max{(S^{d}_m)} \cdot \langle \hat{\mathbf{u}}(\text{e}^{q}_{m}) , \hat{\mathbf{u}}(\text{e}^{d}_{m}) \rangle \nonumber\\
    &= \sum_{m=0}^{M-1}  \max{(S^{q}_m)} \cdot \max{(S^{d}_m)} \mathds{1}_{\{ \text{e}^{q}_{m}= \text{e}^{d}_{m} \}}
\end{align}
\noindent Observing Eq.~(\ref{eq:approximate_dot}), in order to compute the approximate inner product of lexical vectors, each query (passage) can be alternatively represented as two $M$-dimension dense vectors:
\begin{align}
    \qdlr^{\text{val}} &= \left( \max{(S^{q}_{0})},\cdots,\max{(S^{q}_{M-1})} \right) \in \mathbb{R}^{M}  \nonumber \\
    \qdlr^{\text{idx}} &= \left(\text{e}^{q}_{0},\cdots,\text{e}^{q}_{M-1}\right) \in \mathbb{N}^{M} \\
    \ddlr^{\text{val}} &= \left(\max{(S^{d}_{0})},\cdots,\max{(S^{d}_{M-1})}\right) \in \mathbb{R}^{M} \nonumber \\
    \ddlr^{\text{idx}} &= \left(\text{e}^{d}_{0},\cdots,\text{e}^{d}_{M-1}\right) \in \mathbb{N}^{M}, 
\end{align}
where $\qdlr^{\text{val}}$ ($\ddlr^{\text{val}}$) is the dense vector storing the $M$ maximum values from the query (passage) slices, and $\qdlr^{\text{idx}}$ ($\ddlr^{\text{idx}}$) is the integer dense vector storing the entries with the maximum value in the corresponding slices. 

Formally, a dense lexical representation (DLR) is a pair of vectors comprising a ``value'' vector and an ``index'' vector, as indicated in Figure~\ref{fig:idea}.
For a query, the DLR is $(\qdlr^{\text{val}}, \qdlr^{\text{idx}})$ and for each passage, $(\ddlr^{\text{val}}, \ddlr^{\text{idx}})$. 
Intuitively, the ``value'' vector stores the most important term weight in each slice while the ``index'' vector stores the position of the corresponding terms in each slice. 
Using DLRs, Eq.~(\ref{eq:approximate_dot}) can be rewritten as:
\begin{align}
\label{eq:approximate_dot1}
    \langle\qbow, \dbow\rangle \approx \sum_{m=0}^{M-1}   \qdlr^{\text{val}}[m] \cdot \ddlr^{\text{val}}[m] \cdot \mathds{1}_{\{ \qdlr^{\text{idx}}[m]= \ddlr^{\text{idx}}[m] \}},
\end{align}
where $\qdlr^{\text{val}}[m]$ ($\ddlr^{\text{val}}[m]$) and $\qdlr^{\text{idx}}[m]$ ($\ddlr^{\text{idx}}[m]$) is the $m$-th entry of the query (passage) DLR. 
Note that the query (passage) DLR represents approximations of the original lexical query (passage) vector, $\qbow$ ($\dbow$).
Thus, the computation in Eq.~(\ref{eq:approximate_dot1}) using DLRs is an approximation of the original inner product between the lexical vectors. 

To simplify Eq.~(\ref{eq:approximate_dot1}), we define a new operation, \textit{gated inner product} (GIP) as follows: 
\begin{align}
\text{GIP}(\mathbf{q}, \mathbf{d}, \mathbf{g}) \triangleq \sum_{m=0}^{M-1} \mathbf{q}[m]\cdot \mathbf{d}[m] \cdot \mathbf{g}[m],
\end{align}
where $\mathbf{g}$ is the gate vector with entries either equal to 0 or 1 (i.e., $\mathbf{g}[m] \in \{0,1\}$). 
Thus, we can consider Eq.~(\ref{eq:approximate_dot1}) as a GIP operation between query $(\qdlr^{\text{val}}, \qdlr^{\text{idx}})$ and passage $(\ddlr^{\text{val}}, \ddlr^{\text{idx}})$ DLRs:
\begin{align}
\label{eq:dlr_gip}
 \langle\qbow, \dbow\rangle \approx \text{GIP}(\qdlr^{\text{val}}, \ddlr^{\text{val}}, \gdlr),
\end{align}
where the gate vector $\gdlr[m]=\mathds{1}_{\{ \qdlr^{\text{idx}}[m]= \ddlr^{\text{idx}}[m] \}}$ can be interpreted as the result of lexical matching between two DLRs. 
This is why lexical matching can still be performed in low-dimensional DLRs and is the key difference between GIP and standard inner product.
In summary, DLRs and GIP (representing the original vectors and the inner product, respectively) capture the representation and scoring function in our framework.

\subsection{Independent Model Fusion}

In order to perform retrieval based on the fusion of lexical and semantic representations, we can compute their fusion scores as follows: 
\begin{align}
\label{eq:combine}
 \textrm{sim}_{\textrm{\tiny hybrid}}(q,d) 
  &\triangleq \langle\qbow, \dbow\rangle + {\lambda} \cdot \langle \qcls, \dcls \rangle, \end{align}
where $\lambda$ is a hyperparameter.
Here, $\qcls$ and $\dcls$ can refer to any dense semantic representation, including ``off-the-shelf'' ones such as ANCE, DPR, TAS-B, etc.
In standard implementations, these inner products are computed using completely different systems, for example, using Lucene and Faiss for the first and second terms, respectively.
Our work aims to avoid this via a unified framework.

We approximate the first term in Eq.~(\ref{eq:combine}) using Eq.~(\ref{eq:dlr_gip}) and rewrite the second term as a special case of GIP when all the entries in the gate vector are equal to one:
\begin{align}
\label{eq:combine_approx}
 \textrm{sim}_{\textrm{\tiny hybrid}}(q,d) &\approx \text{GIP}(\qdlr^{\text{val}}, \ddlr^{\text{val}}, \gdlr)+ {\lambda} \cdot \text{GIP}( \qcls, \dcls , \mathbf{1}) \nonumber \\
 &= \text{GIP}(\underbrace{\qdlr^{\text{val}}\oplus \sqrt{\lambda} \cdot \qcls}_{\qdhr}, \underbrace{\ddlr^{\text{val}}\oplus \sqrt{\lambda} \cdot \dcls}_{\ddhr}, \underbrace{\gdlr\oplus \mathbf{1}}_{\gdhr}),
\end{align}
where $\mathbf{1}$ is a vector of all ones with the same dimension as [CLS] and $\oplus$ is vector concatenation. 
Observing Eq.~(\ref{eq:combine_approx}), the fusion score can be considered a GIP operation between ($\qdhr^{\text{val}}$,$\qdhr^{\text{idx}}$) and ($\ddhr^{\text{val}}$,$\ddhr^{\text{idx}}$):
\begin{align}
    \qdhr^{\text{val}} &= \qdlr^{\text{val}} \oplus \sqrt{\lambda} \cdot \qcls;  \qdhr^{\text{idx}} = \qdlr^{\text{idx}} \oplus \mathbf{1},  \\
    \ddhr^{\text{val}} &= \ddlr^{\text{val}} \oplus \sqrt{\lambda} \cdot \dcls;  \ddhr^{\text{idx}} = \ddlr^{\text{idx}} \oplus \mathbf{1}.
\end{align}
We call the pair of vectors, $\qdhr^{\text{val}}$ and $\qdhr^{\text{idx}}$ ($\ddhr^{\text{val}}$ and $\ddhr^{\text{idx}}$), query (passage) dense hybrid representations (DHRs).  
Note that DHRs represent a special case of DLRs since they have the same form of representation and scoring function.

\begin{figure}[t]
    \centering
    \resizebox{0.6\columnwidth}{!}{
        \includegraphics{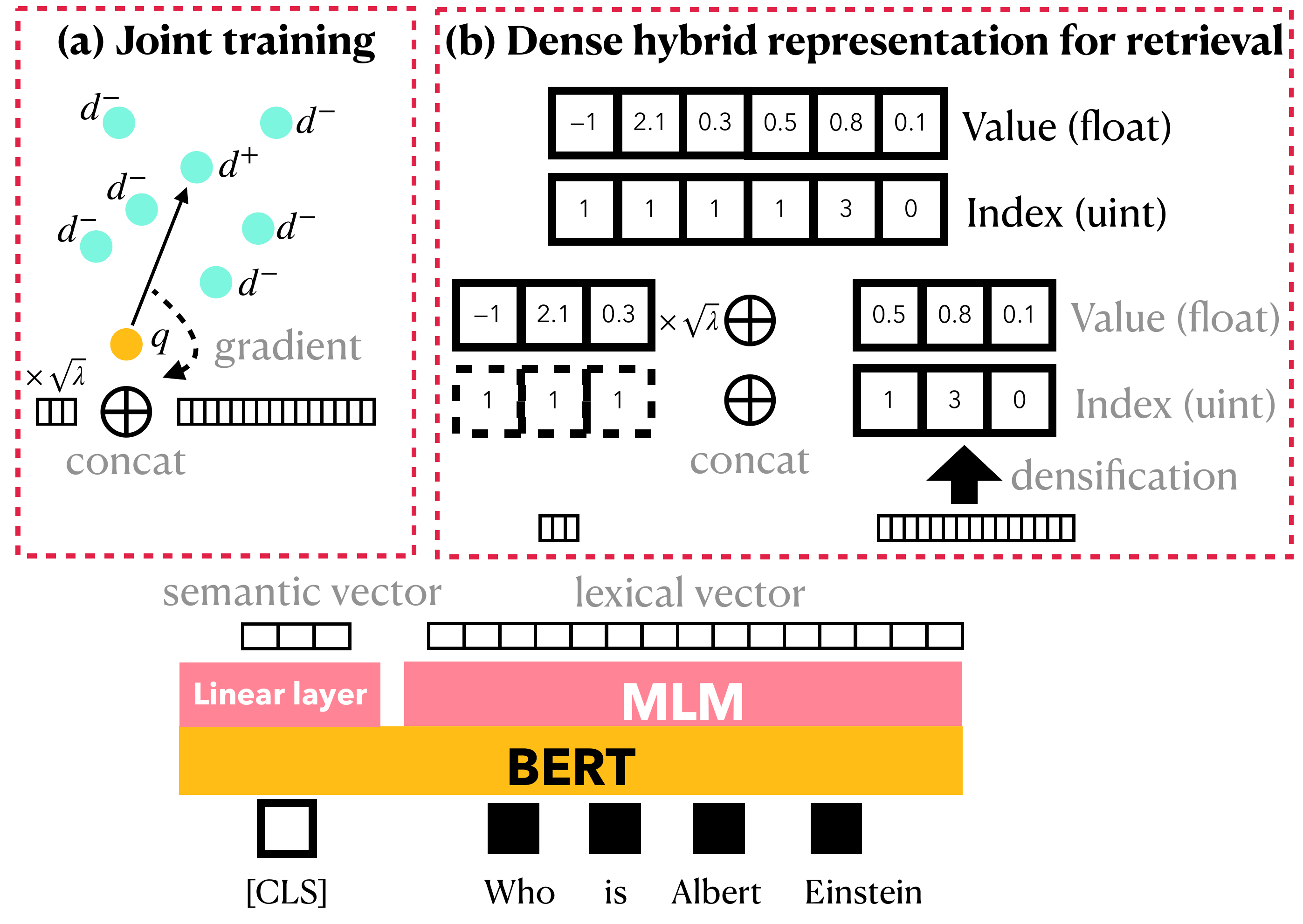}
    }
    \caption{Illustration of single model fusion. (a) During training, we directly concatenate semantic and lexical vectors to compute the relevance score between a query and a passage. (b) During retrieval, we concatenate the semantic and densified lexical vectors to form query and passage DHRs for end-to-end retrieval.
    }
    \label{fig:dhr_illustration}
\end{figure}

\subsection{Single Model Fusion}
\label{subsec:single_model_fusion}

Note that Eq.~(\ref{eq:combine}) is completely general and can be used to combine any arbitrary ``off-the-shelf'' lexical representation (e.g., uniCOIL, SPLADE, etc.) and semantic representation (e.g., ANCE, TAS-B, etc.) for hybrid retrieval.
This approach can be described as fusion of {\it independent} models.
We further study whether our framework can benefit from a single model; thus, we propose to {\it jointly} train lexical and semantic representations within a single model. 
Inspired by \citet{clear}, our intuition is that such learned representations can better complement each other to perform both lexical and semantic matching.

Specifically, given a query $q$, its relevant passage $d^+$ and a set of negative passages $\{d^-_1, d^-_2, \cdots, d^-_l \}$, we train our model by minimizing negative log likelihood of the positive $\{q,d^+\}$ pair over all the passages: 
\begin{align}
\label{eq:nnl}
   \text{NLL} = -\log \frac{e^{\textrm{sim}_{\textrm{\tiny hybrid}}(q,d^+)}}{ e^{\textrm{sim}_{\textrm{\tiny hybrid}}(q,d^+)} + \sum_{j=1}^{l} e^{\textrm{sim}_{\textrm{\tiny hybrid}}(q_i,d_j^-)}}. 
\end{align}
Following \citet{dpr}, we also include both negative and positive passages from the other queries in the same batch as the negatives. 
Note that, as depicted in Figure~\ref{fig:dhr_illustration}, we use the exact fusion score $\textrm{sim}_{\textrm{\tiny hybrid}}(\cdot,\cdot)$ in Eq.~(\ref{eq:combine}) for training, where the original $30522$-dimensional lexical representations are used. 
However, when performing retrieval, we use the approximate fusion score in Eq.~(\ref{eq:combine_approx}). 

\subsection{End-to-End Retrieval with DLRs}
\label{subsec:e2e_retieval}

While it is possible to perform end-to-end retrieval for each query DLR through GIP computations against all passage DLRs in the corpus, we can identify one weakness. 
Unlike standard dense representations, GIP requires $4 \cdot M$ operations, which is more than the standard inner product of $M$-dimensional dense vectors, which only requires $2 \cdot M$ operations. 
When conducting brute-force search over corpus $C$, the difference becomes:\ $4 \cdot M \cdot |C|$ > $2 \cdot M \cdot |C|$.  

\begin{figure}[t]
    \centering
    \resizebox{0.8\columnwidth}{!}{
        \includegraphics{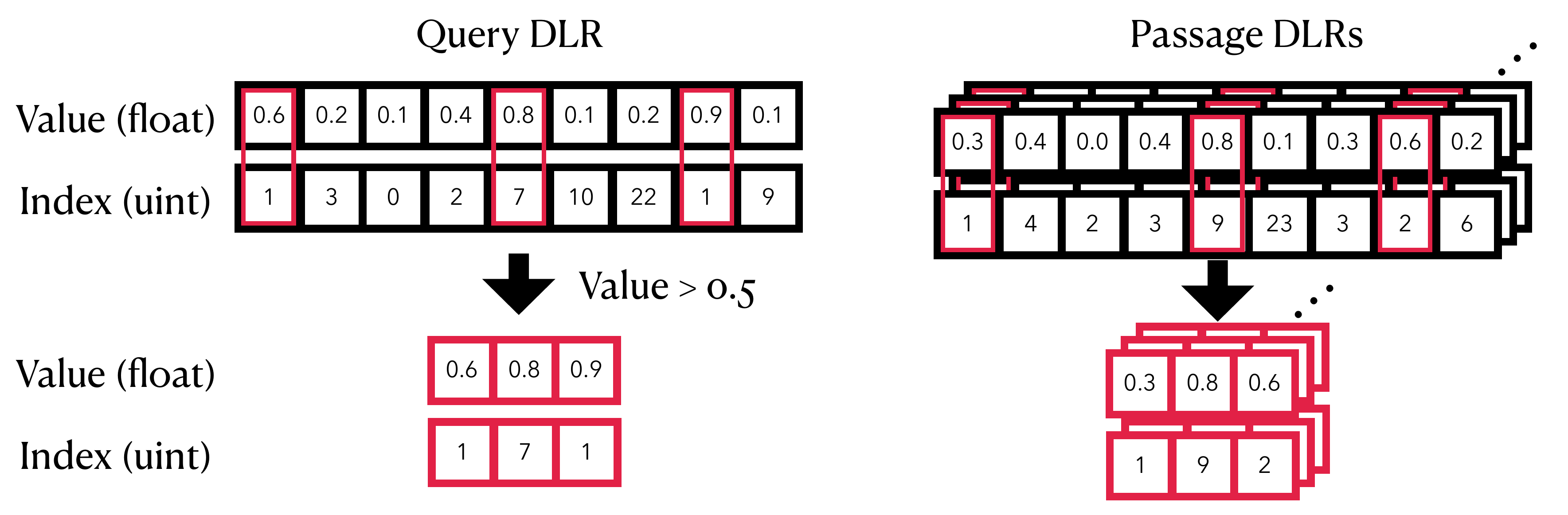}
    }
    \caption{Approximate GIP. DLR retrieval can be approximated by only computing the gated inner product from a few dimensions where the query value vector has values above a threshold (e.g., greater than 0.5 here).
    In two-stage retrieval, we can then precisely rerank the top-$K$ passages using all the dimensions.}
    \label{fig:two-stage}
\end{figure}

To address this issue, we propose a two-stage retrieval approach inspired by previous work~\cite{colbert, bpr}. 
We first retrieve the top-$K$ candidates (where $K \ll |C|$) using approximate score computations and then rerank the $K$ candidates based on the more accurate GIP computations. 
In this work, we propose \textit{approximate} GIP for first-stage approximate retrieval. 
User queries usually contain only a few key terms, which means that when searching the entire corpus, we can perform GIP based on only a few dimensions of DLRs:
\begin{align}
\label{eq:approximate_retrieval}
  \ \ \ \ \ \text{GIP}(\qdlr^{\text{val}}, \ddlr^{\text{val}}, \gdlr) \approx \sum_{m \in \mathcal{M}}  \qdlr^{\text{val}}[m] \cdot \ddlr^{\text{val}}[m] \gdlr[m],
\end{align}
where $\mathcal{M} = \{ m |\qdlr^{\text{val}}[m] > \theta \}$ is the set of indices for the GIP computation, and $\theta$ is a hyperparameter. 
This first-stage retrieval relies on the dimensions where $\qdlr^{\text{val}}[m]$ is above a threshold, as depicted in Figure~\ref{fig:two-stage}.

Note that this approach can also be applied to DHRs, which inherit all their properties from DLRs (and thus can be applied to representations that combine lexical and semantic components). 
In the subsequent main experiments, we use this retrieve-and-rerank approach with approximate GIP as our first-stage retriever. 
See Section~\ref{subsec:study_dlr_retrieval} for further analyses.

\subsection{Choice of Lexical Representation Models}
\label{subsec:model_choice}

Although our approach can be applied to any off-the-shelf model for lexical matching, as described in Section~\ref{subsec:performance_of_DLRs}, in practice, many of the models discussed in Section~\ref{sec:backgroud} still have limitations. 
For example, uniCOIL~\cite{unicoil} and DeepImpact~\cite{deepimpact} require another model for passage expansion, which incurs additional costs for training and inference. 
Thus, in this paper, we choose SPLADE~\cite{splade} as the basis of our lexical representation model, which addresses the above issue by directly learning term expansions and term weights together. 
However, \citet{splade} demonstrate that additional steps are required for tuning a good efficiency--effectiveness tradeoff using a sparsity regularization term~\cite{flop_loss} to enable retrieval with inverted indexes. 
To address this shortcoming, we propose a variant:\ playing off the name of SPLADE, which stands for \textbf{SP}arse \textbf{L}exical \textbf{A}n\textbf{D} \textbf{E}xpansion, we call this variant the \textbf{De}nse \textbf{L}exical \textbf{A}n\textbf{D} \textbf{E}xpansion (DeLADE) model. 
Instead of regularizing vector sparsity, we increase vector density---the exact opposite.
There are two advantages of this design:\ (1) hyperparameter tuning is not required for model training; 
(2) the dense vectors are more robust to our densification approach, as we show in Section~\ref{subsec:performance_of_DLRs}. 

\paragraph{SPLADE} 
Following SparTerm~\cite{sparterm}, SPLADE generates lexical vectors based on the logits of the BERT pretrained masked language model (MLM). 
Consider an input (query or passage) wordpiece sequence $S = (s_1, s_2, \cdots, s_n)$ and its corresponding contexualized token embeddings $H = (\mathbf{h_1}, \mathbf{h_2}, \cdots, \mathbf{h_n})\in \mathbb{R}^{n \times 768}$. 
The logit of each token $s_i$ is computed as follows: 
\begin{align}
    \mathbf{logit}_i = {\text{transform}(\mathbf{h_i})}^{T} \cdot E_{\text{MLM}} + b_{\text{MLM}} \in \mathbb{R}^{|V_{\text{BERT}}|},
\end{align}
where $\text{transform}(\cdot)$ is a linear layer with ReLU activation and LayerNorm, $E_{\text{MLM}} \in \mathbb{R}^{|V_{\text{BERT}}| \times 768}$ is the BERT embedding table, and $b_{\text{MLM}}$ is the bias. 
$V_{\text{BERT}}$ is the vocabulary of BERT wordpiece tokens with size $|V_{\text{BERT}}|=30522$. 
SPLADE generates a single-vector embedding $\mathbf{sp}_{\textrm{\tiny BoW}} \in \mathbb{R}^{|V_{\text{BERT}}|}$ by max pooling over sequence token logits. 
\begin{align}
\label{eq:splade_vector}
    \mathbf{sp}_{\textrm{\tiny BoW}}[v] = \max_{i=1,2,\cdots,n} \log\left(1+\text{ReLU}({\mathbf{logit}}_i[v])\right),
\end{align}
where $\mathbf{sp}_{\textrm{\tiny BoW}}[v]$ is the vector value of the index (or vocabulary ID) $v \in [0,|V_{\text{BERT}}|)$ and $\text{ReLU}(\cdot)$ is the activation function. 
This design naturally promotes vector sparsity since $\log\left(1+\text{ReLU}(\cdot)\right)$ becomes zero for all negative input. 
Together with FLOP regularization loss within a mini-batch $\mathcal{B}$:
\begin{align}
\label{eq:flop}
\mathcal{L}_{\text{FLOP}} = \sum_{v=1}^{|V_{\text{BERT}}|} \left(\frac{1}{\mathcal{B}}\sum_{i=1}^{\mathcal{B}} \mathbf{sp}_{\textrm{\tiny BoW}}[v]\right)^2, 
\end{align}
the sparsity of $\mathbf{sp}_{\textrm{\tiny BoW}}$ can be further increased. 
The overall training loss for SPLADE becomes:
\begin{align}
\mathcal{L} = \text{NLL} + \lambda_{q} \cdot \mathcal{L}^q_{\text{FLOP}} + \lambda_{d} \cdot \mathcal{L}^d_{\text{FLOP}}.
\end{align}
Thus, SPLADE training requires tuning two hyperparameters, $\lambda_{q}$ and $\lambda_{d}$, in order to obtain a good effectiveness--efficiency tradeoff. 
We refer readers to previous work~\cite{flop_loss, splade} for more details. 

\paragraph{DeLADE} 
We make a slight revision to Eq.~(\ref{eq:splade_vector}) by replacing the activation function $\text{ReLU}(\cdot)$ with $\text{softmax}(\cdot)$ to promote vector density. 
\begin{align}
\label{eq:delade_vector}
    \mathbf{ds}_{\textrm{\tiny BoW}}[v] = \max_{i=1,2,\cdots,n} w_i \cdot  \text{softmax} ({\mathbf{logit}}_i)[v],
\end{align}
where $w_i = \mathbf{h}_i^T W + b$ is the linear transformation of $\mathbf{h}_i$ as the term weight (i.e., capturing term importance) of token $s_i$; $\text{softmax} ({\mathbf{logit}}_i)$ can be interpreted as the contextualized representation of token $s_i$, as a probability distribution over $V_{\text{BERT}}$. 
The intuition behind this design is that a sequence is represented as the max pooling of all contextualized representations, while giving more weight to important tokens.  
Furthermore, in contrast to ReLU, softmax is the activation function used in BERT MLM pretraining, which also ensures that the output vector is dense.      
Since vector sparsity is not a concern for retrieval latency in our framework, DeLADE can focus on optimizing negative log likelihood loss without tuning additional hyperparameters to balance efficiency considerations. 

\section{Experimental Setup}

\subsection{Dataset Descriptions}

\paragraph{In-domain IR datasets.}
We use the MS MARCO passage ranking dataset introduced by \citet{marco}, comprising a corpus of 8.8M web passages with the following public query sets for evaluation:\ 
(a) MS MARCO dev:\ 6980 queries comprise the development set for the MS MARCO passage leaderboard, with on average one relevant passage per query.
Following established procedure, we report MRR@10 and R@1000 as the evaluation metrics.
(b) TREC DL~\cite{trec19dl, trec20dl}:\ the organizers of the 2019 (2020) Deep Learning Track at the Text REtrieval Conference (TREC) released 43 (53) queries with graded relevance labels, where (query, passage) pairs were annotated by NIST assessors.
We report nDCG@10 for these two evaluation sets.

\paragraph{Out-of-domain IR datasets.}
We use BEIR, recently introduced by \citet{beir}, which contains 18 distinct IR datasets spanning diverse domains and tasks, including retrieval, question answering, fact checking, question paraphrasing, and citation prediction. 
Each individual dataset comprises its own corpus, queries, and relevance judgements.
Following previous work~\cite{splade-v2, colbert-v2}, we conduct zero-shot retrieval on 13 of the 18 datasets.
Model retrieval effectiveness is evaluated in terms of nDCG@10 and R@100, except for TREC-COVID, where we follow \citet{beir} and use ``capped'' Recall@100 instead of ``regular'' R@100.

\subsection{Models}

\paragraph{Lexical retrieval models.}
To evaluate our approach to densifying lexical representations, we conduct experiments on four lexical matching models:\ two models based on whole word matching, with large vocabulary sizes (millions of distinct terms), and three models that operate on the wordpiece vocabulary of BERT (30522 distinct tokens), including our proposed DeLADE model. 

The whole word matching approaches in more detail:\ 

\begin{enumerate}
\item BM25:\ We can characterize this classic retrieval model as generating heuristically assigned term weights. 

\item DeepImpact~\cite{deepimpact}:\ Term expansion is first applied to each passage in the collection using doc2query--T5~\cite{doctttttquery}.
Then, the encoder (a two-layer MLP with ReLU activation) projects token embeddings from the final layer of BERT into contextualized term weights for each input token in the query and expanded passage. 

\end{enumerate}

\noindent For BM25 and DeepImpact, we output term weights for each query and passage using Pyserini~\cite{pyserini} and then randomly assign each unique term to a unique vocabulary ID to form a high-dimensional sparse vector.

We describe the wordpiece token matching approaches in more detail: 

\begin{enumerate}
\item uniCOIL~\cite{unicoil}:\ This model is similar to DeepImpact, but one main difference is that uniCOIL performs lexical matching on BERT wordpiece tokens during both training and retrieval. 

\item SPLADE~\cite{splade-v2}: To be precise, we refer to the SPLADE-max model, which uses BERT pretrained MLM to project query (or passage) tokens into a $|V_{\text{BERT}}|$-dimensional sparse lexical representation, where $|V_{\text{BERT}}|=30522$ is the vocabulary size of BERT wordpiece tokens. 

\item DeLADE: Our proposed variant of SPLADE encodes queries and passages into $|V_{\text{BERT}}|$-dimensional dense lexical representations. 
\end{enumerate}

For uniCOIL\footnote{\url{https://huggingface.co/castorini/unicoil-msmarco-passage}} and SPLADE,\footnote{\url{https://github.com/naver/splade/tree/main/weights/splade_max}} we use checkpoints provided by the authors to generate vector representations for each query and passage. 
We refer to the DLRs from a base model as $\text{model}_\text{\tiny DLR}(\text{dim})$, where dim denotes the dimensionality of the DLR.
For example, $\unicoildlr(768)$ refers to DLRs of 768 dimensions using uniCOIL as the base model.

\paragraph{Densification.} 
We densify high-dimensional representations by first removing some dimensions to make the dimensionality a multiple of 768; then, we further divide each vector into 768 slices. 
For example, the 30522-dimensional representations are densified into 768-dimensional vectors by (1) discarding the first 570 unused tokens in the BERT vocabulary; (2) dividing the remaining 29952 tokens into 768 slices.
Each of the slices contains 39 distinct vocabulary items; that is, $M=768$ and $N=39$.
In our main experiments, we use the stride slicing strategy, although in Section~\ref{subsec:performance_of_DLRs} we examine the effects of alternatives. 
In addition, we conduct experiments to densify into 256 and 128 dimensions, where there are 117 and 234 tokens in each slice, respectively. 
The ``value'' and ``index'' dense vectors (see Figure~\ref{fig:idea}) are stored as \texttt{float16} and \texttt{uint8}, respectively. 
Note that since we randomly assign a vocabulary ID to each word for BM25 and DeepImpact, the discarded words are also randomly chosen. 
In addition, \texttt{uint8} is sufficient to represent the index vectors for the lexical models that rely on  wordpiece token matching, while \texttt{uint16} is required for BM25 and DeepImpact. 
The detailed settings for densifying the vectors from different models are shown in Table~\ref{tb:densify_setting}. 
In addition, we also list the standard dense [CLS] vector storage setting under our framework for comparison.

\begin{table}[t]
	\caption{Comparison of Vector Densification Settings for the MS MARCO Passage Corpus
	}
	\label{tb:densify_setting}
	\centering
\resizebox{0.8\width}{!}{  
    \begin{threeparttable}
    \begin{tabular}{l!{\color{black}\vrule}rccc}
	\toprule
 Model & vocabulary size &  discarded vocabulary  &  value vector type & index vector type\\
 \midrule
 BM25 & 2,660,824 & 472 & \texttt{float16} & \texttt{uint16}\\
 DeepImpact & 3,514,102 & 502 & \texttt{float16} & \texttt{uint16}\\
 uniCOIL/SPLADE/DeLADE & 30,522  & 570& \texttt{float16} & \texttt{uint8}\\
 Dense $\text{[CLS]}$ & -  & -& \texttt{float16} & \tnote{$\star$}\\
         \bottomrule
	\end{tabular}
    \begin{tablenotes}
	\item[$\star$] Note that we do not have to store the index vector for [CLS] since it is a vector of all ones.
    \end{tablenotes}
    \end{threeparttable}
	}
\end{table}

\paragraph{Single model fusion.} 
In practice, we can jointly train the [CLS] representation with any of the above lexical retrieval models using a single BERT model. 
However, as discussed in Section~\ref{subsec:model_choice}, DeepImpact and uniCOIL require an additional model for passage expansion, and thus they are not ideal base models. 
For SPLADE, FLOP regularization and hyperparameter tuning are required to achieve a good effectiveness--efficiency balance, also making it not an ideal base model.
Instead, we choose our proposed DeLADE model as the base lexical model and refer to the single fusion model as $\text{DeLADE+[CLS]}$.
The corresponding DHR is called $\deladedhr(\text{dim})$, where $\text{dim}$ denotes the dimensionality of the lexical component, DLR. 
We project the [CLS] vectors into 128 dimensions with a linear layer. 

\paragraph{Training and inference details.}
Our proposed models, DeLADE and $\text{(DeLADE+[CLS])}$, are implemented using Tevatron~\cite{tevatron} and trained on a single Tesla V100 GPU with 32 GB memory.
We train our models using \texttt{distilbert-base-uncased}~\cite{distilbert} for 6 epochs (around 100k steps) with learning rate 7e-6. 
Each batch includes 24 samples, and for each query, we randomly sample one positive and seven negative passages. 
All the negatives are sampled from the MS MARCO ``small'' triples training set, which is created using BM25. 
We set the maximum input length for the query and the passage to 32 and 150 (including the special tokens [CLS] and [SEP]), respectively, at both training and inference stages, except for the BEIR dataset, where we set the maximum input length to 512 for both the query and the passage at inference time. 
For $\text{(DeLADE+[CLS])}$, we set $\lambda$ to one during training and inference. 

\paragraph{Advanced training techniques.}
To further compare with other state-of-the-art retrievers, we also train our model with knowledge distillation (KD)~\cite{distilling_hinton} and hard negative mining (HNM), denoted $\text{(DeLADE+[CLS])}^{+}$. 
Specifically, we use $\deladedhr(128)$ to retrieve the top-200 passages using the 8M queries in the training set and then retrain $\text{(DeLADE+[CLS])}$ using ColBERT as the teacher model. 
Note that our ColBERT model (initialized from \texttt{distilbert-base-uncased}) is trained with the soft labels provided by \citet{margin_mse}.\footnote{\url{https://github.com/sebastian-hofstaetter/neural-ranking-kd}}  
Following \citet{tct}, we use KL divergence as the listwise KD loss, which considers all the in-batch negatives. 
For each batch, we include 288 triples (i.e., a query with positive and negative examples) by randomly sampling negatives from the top-200 hard negatives. 
We train $\text{(DeLADE+[CLS])}^{+}$ on three Tesla V100 GPUs.  

\subsection{Retrieval Implementation and Settings}
\label{subsec: retrieval_implementation_and_settings}

Operations in our dense representation framework can be implemented by existing packages that support common array operations, which makes our approach easy to implement and to deploy in real-world settings.
Specifically, our DLR and DHR retrieval experiments are performed using a custom PyTorch implementation, which means that training and retrieval experiments can be conducted within the same execution environment (on GPUs).
Our experiments can be performed without an additional toolkit for nearest neighbor search such as Faiss~\cite{faiss}.

For experimental results reported in Section~\ref{subsec:performance_of_DLRs}, we use two-stage retrieval with the following settings, $\theta=\{1,1,1,1,0.1\}$, for the corresponding lexical models, \{BM25, DeepImpact, uniCOIL, SPLADE, DeLADE\}. 
For experimental results reported in Section~\ref{subsec:independent_model_fusion_result} and Section~\ref{subsec:single_model_fusion_result}, we use two-stage retrieval with $\theta=0.3$ for the independent fusion models, i.e., $(\text{BM25+ANCE})_{\text{\tiny DHR}}$ and $(\text{uniCOIL+ANCE})_{\text{\tiny DHR}}$, and single model fusion, i.e., $\text{(DeLADE+[CLS])}_{\text{\tiny DHR}}$ and $\text{(DeLADE+[CLS])}^{+}_{\text{\tiny DHR}}$. 
We set $K=10000$ in all experiments.

As already noted, our DLR and DHR retrieval experiments are conducted in PyTorch directly.
For the other dense retrieval models, we use Faiss \texttt{FlatIP} GPU indexes as points of comparison. 
We perform all retrieval experiments using a single NVIDIA RTX A6000 with batch size one. 
For retrieval using inverted indexes, we use Pyserini (which is built on Lucene) and measure retrieval latency using a single thread on a Linux machine with two 2.1 GHz Intel Xeon Platinum 8160 CPUs and 944G of RAM.
Note that in our main experiments, we primarily compare retrieval latency between different types of text representations (i.e., sparse lexical, dense semantic, and DLR/DHR). 
Thus, we exclude the encoding time of the query text from the neural models for simplicity. 
Query encoding latency on a GPU is around 10--20 ms per query for the backbones (\texttt{bert-base-uncased} or \texttt{distilbert-base-uncased}) used by the compared neural models (except for {GTR\tiny XXL}~\cite{gtr}); thus, this does not have much impact on the latency comparison between neural models.
We refer readers to Table~\ref{tb:online_latency_measurement} for detailed online retrieval latency measurements of our models on the CPU and GPU.

\section{Results}

In this section, we present our experimental results and discuss each research question in turn. 

\subsection{Quality of DLR Approximations}
\label{subsec:performance_of_DLRs}

To begin, we densify lexical representations into DLRs of different dimensions to investigate our first research question:
\RQone

\input{dlr_comparison}

\smallskip \noindent
Table~\ref{tb:dim_reduction} shows the results of densifying different lexical representations.
In these experiments, we use stride slicing, but examine the impact of different slicing strategies below.
The first row in each block reports the retrieval effectiveness of the original lexical representations using inverted indexes, which can be considered an upper bound. 
We also report the effectiveness difference (shown as a percentage) between each method and the upper bound using inverted indexes.
Note that since DeLADE is trained without any sparsity constraints, the output query and passage representations are quite dense and hence impractical for retrieval using inverted indexes.
For this condition, we use $\texttt{faiss.FlatIP}$ brute-force search instead. 
Index size is provided only as a reference, and we omit query latency since it is not comparable to retrieval with inverted indexes.
We also report the number of tokens per passage (i.e., vector indices with non-zero weights) for each vector densification condition.

\input{compression_comparison.tex}

Our method is able to densify high-dimensional lexical vectors into 768-dimensional vectors with only a small retrieval effectiveness drop. 
As the vectors are further densified into smaller dimensions, retrieval effectiveness drops more, which can be explained by information loss since the number of tokens are reduced through max pooling in each slice; i.e., collision between tokens appearing in the same slice. 
Compared to models that use whole word matching, BM25 and DeepImpact, wordpiece matching models appear to be more robust to vector densification. 
For example, with 128-dimensional DLRs, uniCOIL sees only $4.6\%$ and $1.5\%$ degradation in MRR@10 and R@1K, respectively, while BM25 sees $10\%$ and $4.9\%$ degradation in MRR@10 and R@1K, respectively. 
In contrast to whole word matching models, wordpiece matching models represent many terms with multiple wordpiece tokens; thus, effectiveness is less sensitive to vector densification since there is more redundancy in the representation.
In addition, BM25 is the only model that does not benefit from passage expansion; thus, reducing the token space causes greater effectiveness loss.

Among wordpiece matching models, uniCOIL sees only modest retrieval effectiveness degradation for 256 and 128 dimensions while SPLADE sees larger effectiveness drops. 
This is likely because SPLADE represents each passage with more wordpiece tokens than uniCOIL does; thus, there are more collisions as the vectors are densified into smaller dimensions. 
In contrast to SPLADE, although there are more collisions from vector densification with $\deladedlr$, our model sees less retrieval effectiveness degradation. 
We attribute this robustness to collisions to DeLADE's full expansion over the entire wordpiece vocabulary space without any sparsity regularization; that is, there appears to be more ``redundancy'' in the representations.
This result indicates that DeLADE is a better alternative to SPLADE for dense lexical matching.

Finally, a comparison of DLR and SPAR~\cite{spar} shows the advantages of our approach. 
Without any training, our 256-dimensional DLRs applied to BM25 and uniCOIL are able to compete with SPAR. 
Furthermore, our approach consumes less space and exhibits lower retrieval latency. 
Note that the index vectors for the whole word matching models (BM25 and DeepImpact) are stored in \texttt{unit16} due to larger vocabulary sizes; thus, they consume more storage compared to the wordpiece matching models.

In addition, we observe that retrieval latency using inverted indexes is sensitive to the average number of tokens per passage (i.e., vector sparsity). 
For example, a 35\% increase from 68 to 92 (uniCOIL vs.\ SPLADE) leads to more than 60\% latency increase. 
In contrast, DLRs see lower retrieval latency, which appears to be insensitive to the average number of tokens per passage; for example, $\unicoildlr$ and $\deladedlr$ exhibit large differences in vector sparsity but have comparable latency when using vectors with the same dimensions.
Even with different dimensions, our approach does not exhibit much variability in retrieval latency. 
We attribute this advantage to our two-stage retrieval approach, where computationally expensive end-to-end retrieval only relies on a few dimensions (see Section~\ref{subsec:study_dlr_retrieval} for more details).

To compare our approach to the unsupervised vector compression techniques shown in Table~\ref{tb:compression_comparison}, we use the FiQA-2018 test collection in BEIR~\cite{beir}, which is based on a medium-size corpus with 57K passages.
Compared to MS MARCO (8.8M passages), it is less computationally demanding to perform unsupervised training for the various vector compression techniques. 
We use DeLADE as the base model and report results (nDCG@10 and R@100 as quality metrics and storage as the efficiency metric) in Table~\ref{tb:unsupervised_dim_reduction}. 
Our experiments follow Faiss instructions\footnote{\url{https://github.com/facebookresearch/faiss/wiki/Faiss-indexes}} for applying the unsupervised vector compression techniques.
The \texttt{FlatIP} index (without any compression) provides the performance upper bound.
For locality-sensitivity hashing (LSH), we use \texttt{faiss.IndexLSH} with $\texttt{nbits}= 8 \cdot 768$; for product quantization (PQ), we use \texttt{faiss.IndexPQ} with $\texttt{nbits}=8$ and $M=768,256,128$.\footnote{We also tried OPQ~\cite{OPQ} training before performing PQ; however, the effectiveness is much worse.}
We observe that \texttt{PQ768} can retain most of the information from the original high-dimensional vectors (less than 10\% retrieval effectiveness drop) while LSH cannot effectively compress even a 30K-dimensional vector into binary codes.
When we further compress the original vectors with \texttt{PQ256} and \texttt{PQ128}, retrieval effectiveness drops more than 10\% (and even more for \texttt{PQ128}). 
In contrast, DLR shows retrieval effectiveness drops less than 10\% and performs consistently better than product quantization with the same storage size.
Furthermore, our approach does not require any supervised or unsupervised training.

\begin{table}[t]
	\caption{Comparison of Different DLR Slicing Strategies on MS MARCO (Dev).}
	\label{tb:ablation}
	\centering
	\resizebox{0.8\width}{!}{  
    \begin{tabular}{lccccccccc}
	\toprule
	&\multicolumn{3}{c}{$\unicoildlr(768)$} &\multicolumn{3}{c}{$\spladedlr(768)$} &\multicolumn{3}{c}{$\deladedlr(768)$} \\
	\cmidrule(lr){2-4} \cmidrule(lr){5-7} \cmidrule(lr){8-10}
	 Strategy & \# tokens/doc &  MRR@10&  R@1K &  \# tokens/doc &  MRR@10&  R@1K &  \# tokens/doc &  MRR@10&  R@1K \\
\midrule
        Contiguous & 39.27 & 0.334 & 0.947 & 68.49 & 0.326 & 0.952 & 768 & 0.332 & 0.949 \\
         Stride & 64.15 & 0.349 & 0.957 &86.33& 0.336 & 0.963 & 768 & 0.345 & 0.953 \\
         Random & 64.32 & 0.349 & 0.957 & 87.91 & 0.336 & 0.963 & 768 & 0.345 & 0.953 \\
	\bottomrule
	\end{tabular}
	}
\end{table}

Finally, we examine three different slicing strategies on the wordpiece matching models.\footnote{Recall that the vocabulary IDs for the whole word matching models are randomly assigned; thus, the slicing strategy is considered random.}
We densify lexical representations with 768 dimensions (i.e., $M=768$). 
Table~\ref{tb:ablation} reports results for the three lexical retrieval models, along with the number of tokens per passage for each condition. 
We notice that stride slicing has the same effectiveness as randomized slicing (only minor differences observed beyond four digits).
However, surprisingly, the contiguous slicing strategy shows degradation in ranking effectiveness and the number of tokens per passage for this condition is smaller than the other slicing strategies. 
This indicates that BERT wordpiece tokens with adjacent token IDs may co-occur with higher probability than any two randomly chosen tokens.
Thus, max pooling over contiguous slices of BERT wordpiece token IDs leads to more collisions compared to the other strategies.
Based on this analysis, we use stride slicing as our default setting in the rest of our experiments.

\subsection{Evaluation of Independent Model Fusion}
\label{subsec:independent_model_fusion_result}

In this section, we describe experiments on fusing different ``off-the-shelf'' lexical and semantic retrieval models and compare their effectiveness and efficiency to other hybrid retrieval methods to answer the following research question:
\RQtwo 

\smallskip \noindent Following \citet{spar}, we conduct experiments on BM25--ANCE and uniCOIL--ANCE fusion on the MS MARCO dev set, reported in rows (1)--(3) and rows (4)--(6), respectively, in Table~\ref{tb:fusion_retrieval_comparison}. 
The first entry in each main block, rows (1) and (4), represents the linear combination approach used in previous work~\cite{dpr, tct, unicoil}, which requires two separate indexes and additional post-processing of the ranked lists. 
For these experiments, we measure latency in Pyserini and report a theoretically optimized system as the maximum latency between Lucene and Faiss retrieval plus 3 ms of post-processing time.
We leave aside the engineering challenge of synchronizing CPU and GPU search necessary to achieve this performance under real-world conditions. 
For SPAR~\cite{spar}, rows (2) and (5), we directly report the ranking effectiveness from the paper and measure the retrieval latency of the Faiss FlatIP index in our environment.
Note that their approach distills uniCOIL's lexical representations into semantic representations and then concatenates them to ANCE for dense retrieval. 
Thus, dimensionality of the representation vectors is reported as 2 $\times$ 768. 
For DHRs, rows (3) and (6), we tune $\lambda$ on MRR@10 using a subset of 100 queries in the training set and set $\theta=0.3$ for two-stage retrieval.

\input{dhr_comparison.tex}

Overall, all three systems yield similar retrieval effectiveness but our DHRs achieve lower retrieval latency. 
Note that our model variant with 768-dimensional lexical and semantic vectors is faster than SPAR using Faiss GPU. 
We attribute this improvement to our two-stage retrieval approach (see Section~\ref{subsec:study_dlr_retrieval} for more details). 
It is worth noting that with a negligible effectiveness drop we can further compress the lexical representations to 128 dimensions. 
Furthermore, we can convert lexical representations into DHRs of any width according to user design requirements {\it without any model retraining}.
This flexibility is one major advantage of our approach.

\input{main_comparison.tex}

\subsection{Evaluation of Single Model Fusion}
\label{subsec:single_model_fusion_result}

With our framework and proposed DeLADE model, fusing lexical and semantic representations becomes easier. 
This motivates us to investigate:
\RQthree

\smallskip \noindent Table~\ref{tb:main_result} compares model performance in terms of retrieval effectiveness and efficiency. 
For efficiency, index size and retrieval latency are measured on the MS MARCO dev set as the point of reference.
The comparison models across the columns are categorized as:\ (1) sparse lexical retrievers, including BM25, docT5q~\cite{doctttttquery}, and SPLADE~\cite{splade-v2}; (2) dense semantic retrievers, including our trained baseline dense retriever (denoted Dense with 768-dimensional \texttt{[CLS]} vectors) and ANCE~\cite{xiong2020approximate}; (3) multi-vector retrievers, including ColBERT~\cite{colbert} and COIL~\cite{coil}. 

For the MS MARCO datasets, we conduct experiments using Pyserini~\cite{pyserini} for all models except for ColBERT\footnote{We copy numbers from \citet{deepimpact}.} and COIL.\footnote{We run COIL using the inference code from the authors' repo at \url{https://github.com/luyug/COIL}\label{note1}.} 
For the BEIR datasets, we directly copy numbers from \citet{contriever}, except for SPLADE\footnote{We run SPLADE-max using the inference code from the authors' repo at \url{https://github.com/naver/splade}.} and COIL.\footref{note1} 
To determine the statistical significance of our results, we perform paired $t$-tests ($p <0.05$) comparing all models except for ColBERT on the MS MARCO datasets. 
For a fair comparison, Table~\ref{tb:main_result} only includes models that use the same baseline training strategy as ours.
Thus, we exclude approaches that depend on other models for expansion~\cite{deepimpact, unicoil, tildev2}, costly training techniques such as knowledge distillation~\cite{rocketqa, tasb, gpl, colberter, colbert-v2, splade-v2}, or special pretraining~\cite{condenser, contriever, gtr}, although see Table~\ref{tb:advanced_comparison} for additional comparisons. 

In terms of our proposed models, we report results on three $\deladedhr$ variants by densifying the lexical components into 128, 256, and 768 dimensions. 
The $\deladedhr$ 128- and 256-dimensional variants can be considered ``small vectors'', to compare with single-vector (sparse lexical and dense semantic) retrievers, while the $\deladedhr$ 768-dimensional variant can be considered ``large vectors'', to compare with multi-vector retrievers. 
Note that all three variants are derived from the same model. 
Finally, we report the performance of $\deladedlr(768)$, i.e., without the incorporation of the [CLS] vector.

\paragraph{DHRs vs DLRs.}
We first compare $\deladedhr(768)$ with $\deladedlr(768)$, columns (8) and (b), to examine the effectiveness of single model fusion. 
From the results, we see that $\deladedhr(768)$, which incorporates an additional 128-dimensional [CLS] vector, demonstrates significantly better retrieval effectiveness than $\deladedlr(768)$ for both in-domain and out-of-domain datasets. 
Furthermore, $\deladedhr(256)$ outperforms $\deladedlr(768)$ for all metrics, which suggests that incorporating the [CLS] vector can mitigate information loss from our densification approach.  

To further understand how joint training works for $\deladedhr(768)$, we conduct retrieval using the densified 768-dimensional lexical vectors from DeLADE and 128-dimensional \texttt{[CLS]} semantic vectors separately. 
Retrieval effectiveness on the MS MARCO dev set and three BEIR datasets (TREC-COVID, FiQA-2018, and SciFact) is reported in Table~\ref{tb:joint_training_effect}. 
We see that each component is far less effective individually than their hybrid fusion. 
Specifically, the DeLADE and \texttt{[CLS]} components in $\deladedhr(768)$ are less effective than the independently trained $\deladedlr(768)$ and Dense models, respectively; see columns (4) and (8) in Table~\ref{tb:main_result}. 
This result indicates that joint training yields components that are highly complementary, as designed.

To further explore this complementarity of representation, we vary the dimensionality of the semantic component of $\deladedhr(768)$. 
We jointly train $\deladedlr(768)$ with 0, 128, 256, and 768-dimensional [CLS] vectors separately.  
Retrieval effectiveness on the MS MARCO dev set and three BEIR datasets (TREC-COVID, FiQA-2018, and SciFact) is reported in Table~\ref{tb:cls_ablation}. 
We see that fusing a small [CLS] vector (e.g., 128 dimensions) in training improves retrieval effectiveness, indicating that joint training yields complementary lexical and semantic representations within a single model.
This is consistent with the component analysis above.
However, further increasing the dimensionality of the [CLS] vector does not appear to yield obvious advantages. 
This result indicates that a relatively small [CLS] vector is sufficient to complement the lexical component.

\paragraph{DHRs vs single-vector models.}
First, a comparison of single-vector retrievers shows that sparse lexical representations have better storage efficiency than dense semantic representations. 
For example, docT5q, column (2), with an index less than 1 GB, exhibits better out-of-domain retrieval effectiveness than dense semantic retrievers, columns (4) and (5).
Similarly, SPLADE, column (3), outperforms the dense retrievers for both in-domain and out-of-domain conditions with an index of only 2.6 GB. 
However, SPLADE requires over 7 times the retrieval latency of docT5q and other dense retrievers.
As lexical--semantic hybrid representations, DHRs exhibit better retrieval effectiveness with modest retrieval latency and index storage consumption.
We see that the 128- and 256-dimensional variants of $\deladedhr$, columns (9) and (a), outperform the single-vector retrievers, columns (1) to (5), for both in-domain and out-of-domain datasets in terms of retrieval effectiveness, and furthermore achieves lower query latency. 
In addition, the two $\deladedhr$ variants have modest index sizes; for example, the Dense model, column (4), requires 26 GB to store the MS MARCO passage corpus (using a Faiss \texttt{FlatIP} index) while $\deladedhr(128)$ only consumes 5.4 GB.

\paragraph{DHRs vs multi-vector models.}
We compare retrieval models with large vectors, $\deladedhr(768)$ and multi-vector retrieval models, shown in columns (b), (6), and (7). 
Although their in-domain retrieval effectiveness does not appear to be very different, ColBERT falls behind COIL and $\deladedhr(768)$ in out-of-domain evaluation. 
This result suggests that lexical matching remains a key component for generalization. 
Compared to COIL, $\deladedhr(768)$ exhibits comparable retrieval effectiveness but has a much smaller index (60 GB vs 22 GB). 

It is worth mentioning that COIL has other variants with smaller indexes; for example, the configuration with 8-dimensional tokens plus 128-dimensional [CLS] embeddings consumes 14 GB and yields 0.347 (0.956) MRR@10 (R@1K) on the MS MARCO dev queries.\footnote{We refer readers to \citet{coil} for more details.} 
This variant is still slightly less effective than our models, $\deladedhr(256)$ and $\deladedhr(128)$, with index sizes of 8.6 GB and 5.4 GB, respectively. 
Furthermore, all our variants are derived from the same model without retraining. 
This comparison demonstrates the advantages of single model fusion under our proposed dense representation framework. 

\paragraph{Summary.}
We observe that dense retrievers generally perform well in domain, while sparse retrievers appear to yield stronger generalization capabilities. 
Our results demonstrate that DHRs inherit advantages of both lexical and semantic matching. 
Specifically, our $\deladedhr$ model achieves competitive retrieval effectiveness in both evaluation settings with low retrieval latency and modest index storage consumption. 
This advantageous effectiveness--efficiency tradeoff makes the design of single model fusion with DHRs attractive. 
It is also worth noting that $\deladedlr(768)$ outperforms SPLADE slightly,\footnote{Our reproduced numbers are slightly better than \citet{splade-v2} except for T\'{o}uche-2020, where we use v2 instead of v1.} especially in term of nDCG. 
This result shows that DeLADE is a good alternative under our framework compared to SPLADE using inverted indexes in the scenario where lower retrieval latency is more important than index size.

\input{advanced_comparison.tex}

\paragraph{DHRs vs more advanced retrieval models.}
Finally, we compare $\deladedhr^+$, which uses advanced training strategies, to the effectiveness and efficiency of existing state-of-the-art retrievers in Table~\ref{tb:advanced_comparison}. 
For sparse lexical retrieval models, we report results from SPLADEv2~\cite{splade-v2}, which uses two rounds (i.e., training with BM25 and hard negatives) of knowledge distillation from a cross-encoder teacher.

For dense semantic retrieval models, we include three representative models for comparison:\
(1) TAS-B~\cite{tasb} distills knowledge from multiple teachers using a topic-aware negative sampling strategy.
(2) RocketQAv2~\cite{rocketqav2} leverages a student--teacher joint training approach to make the dense retriever better mimic a cross-encoder teacher.
(3) Contriever~\cite{contriever} leverages pretraining by combining advanced contrastive learning techniques with an Inverse Cloze Task (ICT) variant. 
We also include dense retrievers trained with much larger backbone models (i.e., more model parameters):\ 
(1) GPL~\cite{gpl} trains an expert model for each target dataset in BEIR. 
(2) GTR~\cite{gtr} trains even larger encoder models; the authors' T5-3B and T5-11B models correspond to {GTR\tiny XL} and {GTR\tiny XXL}, which also combine pretraining, knowledge distillation (KD), and hard negative mining (HNM).\footnote{GTR is fine-tuned on MS MARCO training queries with hard negatives denoised by a cross-encoder.}
Finally, for multi-vector retrieval models, we report ColBERTv2~\cite{colbert-v2}, which also combines KD and HNM, and at the same time compresses the multi-vectors into a smaller index.

We point out that it is not easy to fairly compare models with more advanced and costly training techniques since, as shown in Table~\ref{tb:advanced_comparison}, there are many substantive differences that cannot be captured by general descriptive labels. 
Even considering a general method such as knowledge distillation, there are many different implementations. 
For example, we use a lightweight ColBERT teacher, while GPL, ColBERTv2, and SPLADEv2 use a more expensive cross-encoder teacher.
While end-to-end results are comparable since the evaluations use the same test collections, it is difficult to attribute effectiveness differences to specific components.

Nevertheless, it is possible to draw some conclusions from these experiments.
We first observe that dense semantic vectors from smaller models, in columns (d)--(f), do not appear to perform well in both in-domain and out-of-domain evaluations. 
For example, Contriever performs well on BEIR but lags behind most models on in-domain evaluation and the reverse trend can be observed for RocketQAv2. 
To improve the generalization capability of dense retrieval models, existing work either leverages multiple expert models for multi-domain datasets as GPL, column (g), or larger pre-trained models as GTR, columns (h)--(i).

On the other hand, SPLADEv2 and ColBERTv2, columns (c) and (j), leverage representations with more expressive capacity than dense semantic vectors and appear to show equally good generalization capability without adding model parameters. 
However, these models sacrifice retrieval efficiency to gain this generalization capability. 
For example, both SPLADEv2 and ColBERTv2 are slower than dense semantic models. 
GTR also sacrifices query encoding latency, which is excluded in our latency measurement.\footnote{{GTR\tiny XL} and {GTR\tiny XXL} query encoding consume 96 and 349 ms, respectively, as opposed to 10 ms for TAS-B, reported by \citet{gtr}}

In contrast, $\deladedhr^+(768)$ not only yields competitive zero-shot retrieval effectiveness on BEIR compared to {GTR\tiny XXL}, ColBERTv2, and SPLADEv2, but also maintains low retrieval latency. 
In addition, our $\deladedhr^+(256)$, column (l), still yields better overall performance (i.e., effectiveness and efficiency) compared to the single-vector dense retrievers, TAS-B, GPL, and Contriever. 
We notice that GTR performs particularly well in some QA datasets (e.g., NQ and FiQA-2018), likely because GTR includes more training data such as QA pairs mined from the web and the NaturalQuestions dataset. 
We also note that $\deladedhr$ cannot compete with RocketQAv2, GTR, and ColBERTv2 on in-domain evaluation since the first two exploit additional QA training pairs and all of them use more expensive cross-encoder teachers, which is orthogonal to our work. 
It is likely that we can further boost $\deladedhr$ effectiveness by leveraging more data and more expensive teachers; however, we leave these explorations for future work.

\subsection{Performance of Two-Stage Retrieval}
\label{subsec:study_dlr_retrieval}

In this section, we further study our approach to end-to-end retrieval with DLRs/DHRs proposed in Section~\ref{subsec:e2e_retieval} to investigate our final research question:
\RQfour

\smallskip \noindent To illustrate how well our proposed approximate GIP operation compares to the more expensive exact GIP operation between DLRs (or DHRs), we use $\deladedlr(768)$ and $\deladedhr(768)$ to conduct end-to-end retrieval experiments on the MS MARCO dev queries.

Figure~\ref{fig:retrieval_comparison} illustrates recall at different cutoffs using approximate GIP with various settings of the parameter $\theta$. 
Exact GIP, which is equivalent to $\theta=0$, is shown as the black dashed line and represents the upper bound; as $\theta$ increases, recall drops, as expected. 
However, the results show that the top-10000 candidates retrieved using approximate GIP include more relevant passages (i.e., has higher recall) than GIP at cutoff 1000. 
That is, the effectiveness drop from approximate (first-stage) retrieval can be remedied by reranking the top-10000 candidates with GIP. 
In addition, we note that $\deladedhr(768)$ suffers a smaller recall drop at the approximate retrieval stage. 
This suggests that [CLS] vectors can help capture relevance and compensate for lower recall from the lexical component with larger values of $\theta$. 
To be clear, $\theta$ is also applied to the [CLS] vector to guide the approximate computations in $\deladedhr$. 
In addition, we also perform retrieval by computing the standard inner product (IP) between two DLRs' or DHRs' value vectors (without considering their index vectors) for comparison. 
We observe that IP, surprisingly, also retrieves more relevant passages at cutoff 10000 than GIP at cutoff 1000, especially for DHRs. 
This indicates that standard inner product between two DLRs' (or DHRs') value vectors can be an alternative to approximate GIP. 
We show below that using IP as first-stage retrieval can benefit DLRs (or DHRs) in end-to-end retrieval on the CPU.

\input{two_stage_figures.tex}

We further examine effectiveness--efficiency tradeoffs after reranking the top-10000 candidates with exact GIP. 
The red lines in Figures~\ref{fig:two-stage_retrieval_comparison}(a) and \ref{fig:two-stage_retrieval_comparison}(b) plot the $\deladedlr(768)$ performance tradeoff curve (ranking effectiveness vs retrieval latency) of approximate GIP and reranking with different values of $\theta$, while the blue dashed line depicts the performance curve of approximate GIP without reranking for comparison.
Retrieval effectiveness with exact GIP (the black triangle) can be considered the upper bound.  
From the blue dashed lines, we observe that approximate GIP substantially reduces retrieval latency but sacrifices effectiveness. 
However, reranking the top-10000 candidates with GIP mostly recovers the effectiveness loss (except for $\theta=0.3$) and requires only an additional 5--10 ms. 

By incorporating the [CLS] vector, the performance tradeoff curves appear to be even better, as shown in Figures~\ref{fig:two-stage_retrieval_comparison}(c) and \ref{fig:two-stage_retrieval_comparison}(d), which are organized in the same manner as Figures~\ref{fig:two-stage_retrieval_comparison}(a) and \ref{fig:two-stage_retrieval_comparison}(b).
For $\deladedhr(768)$, we see no obvious retrieval effectiveness drop, even at $\theta=0.3$. 
This result is consistent with our observation in Figure~\ref{fig:retrieval_comparison} that $\deladedlr(768)$ and $\deladedhr(768)$ retrieve enough relevant passages in the top-10000 candidates using approximate GIP with $\theta \leq 0.3$. 
Finally, we observe that IP only shows a minor R@1K degradation after GIP reranking, indicating that the standard inner product can provide an alternative to approximate GIP. 
This means that end-to-end retrieval with DLRs or DHRs can also be implemented with first-stage IP followed by GIP reranking. 

\input{latency_measurement}

Next, we measure the end-to-end retrieval latency of $\deladedhr$, including both query encoding (including vector densification) and retrieval components, using the same GPU and CPU environments described in Section~\ref{subsec: retrieval_implementation_and_settings}. 
We report the query latency averaged over the 6980 MS MARCO dev queries in Table~\ref{tb:online_latency_measurement}.
With the exception of row (1), for each query, we retrieve the top-10000 passages at the first stage and rerank the candidates using the second stage.

On the GPU, query encoding consumes 12 ms per query. 
With approximate GIP retrieval and GIP reranking, row (2), our lowest end-to-end retrieval latency on the GPU is 45 ms per query. 
With inner product retrieval and GIP reranking, query latency rises slightly to 59 ms per query, row (3).
End-to-end retrieval latency on the CPU can be reduced to less than one second per query by performing inner product first-stage retrieval on the quantized query and passage value vectors, shown in rows (4) and (5). 
Although approximate GIP can improve the latency of first-stage retrieval over 80$\times$ compared to GIP on the CPU, row (2) vs row (1), approximate GIP is still over 2$\times$ slower than inner product, row (2) vs row (3).
This stands in contrast to the GPU, where approximate GIP is {\it faster} than inner product.
We note that approximate GIP requires selection of certain dimensions, i.e., $\mathcal{M}$ in Eq.~(\ref{eq:approximate_retrieval}), from the passage value and index vectors across the entire corpus, and this operation appears to be the latency bottleneck in our implementation using PyTorch on the CPU with a single thread. 
Thus, on the CPU, using the inner product between query and passage vectors is a better choice for first-stage retrieval, which can be further accelerated with approximate nearest neighbor (ANN) search algorithms.

\begin{table*}[t]
	\caption{Effects of $\theta$ on Approximate GIP with the Reconstructed Expanded Query from $\deladedlr(768)$}
\label{tb:showcase}
	\centering
	 \resizebox{0.8\width}{!}{  
	  	 \setlength\tabcolsep{2.5pt}
    \begin{tabular}{c!{\color{lightgray}\vrule}c!{\color{lightgray}\vrule}l}
	\toprule
    \multicolumn{3}{c}{{\bf Original query:} where was the bauhaus built} \\
    \midrule
\multirow{1}{*}{$\theta$} & \multirow{1}{*}{rank} & \multicolumn{1}{c}{Reconstructed (expanded) query terms}\\
 \midrule
 \multirow{3}{*}{0.00} & \multirow{3}{*}{2} & \textcolor{pink}{location}, \textcolor{bleudefrance}{built}, \textcolor{bleudefrance}{\#\#uh}, \textcolor{bleudefrance}{\#\#aus}, \textcolor{bleudefrance}{ba}, \textcolor{pink}{was}, \textcolor{bleudefrance}{house}, \textcolor{pink}{build}, \textcolor{pink}{building}, site, \textcolor{pink}{were}, \textcolor{bleudefrance}{school}, home, church, the,\\
 & &store, originally, \textcolor{pink}{place}, construction, \textcolor{pink}{studio}, \textcolor{pink}{founded}, headquarters, \textcolor{pink}{structure}, later, city, \textcolor{pink}{is},  orig-\\
 & &in, \textcolor{pink}{be}, theater, \textcolor{pink}{college}, first, hotel, \textcolor{pink}{villa}, manufacture \ldots (omit)\\ \arrayrulecolor{lightgray}
  \midrule
  \multirow{1}{*}{0.05} & \multirow{1}{*}{3} & \textcolor{pink}{location}, \textcolor{bleudefrance}{built}, \textcolor{bleudefrance}{\#\#uh}, \textcolor{bleudefrance}{\#\#aus}, \textcolor{bleudefrance}{ba}, \textcolor{pink}{was}, \textcolor{bleudefrance}{house}, \textcolor{pink}{build}, \textcolor{pink}{building}, site, \textcolor{pink}{were}, \textcolor{bleudefrance}{school}\\
  \midrule
  \multirow{1}{*}{0.10} & \multirow{1}{*}{6} & \textcolor{pink}{location}, \textcolor{bleudefrance}{built}, \textcolor{bleudefrance}{\#\#uh}, \textcolor{bleudefrance}{\#\#aus}, \textcolor{bleudefrance}{ba}, \textcolor{pink}{was}, \textcolor{bleudefrance}{house}, \textcolor{pink}{build}, \textcolor{pink}{building}, site\\
   \midrule
  \multirow{1}{*}{0.20} & \multirow{1}{*}{16} & \textcolor{pink}{location}, \textcolor{bleudefrance}{built}, \textcolor{bleudefrance}{\#\#uh}, \textcolor{bleudefrance}{\#\#aus}, \textcolor{bleudefrance}{ba}, \textcolor{pink}{was}, \textcolor{bleudefrance}{house}, \textcolor{pink}{build}\\
   \midrule
   \multirow{1}{*}{0.30} & \multirow{1}{*}{19} & \textcolor{pink}{location}, \textcolor{bleudefrance}{built}, \textcolor{bleudefrance}{\#\#uh}, \textcolor{bleudefrance}{\#\#aus}, \textcolor{bleudefrance}{ba}, \textcolor{pink}{was}, \textcolor{bleudefrance}{house}\\
    \arrayrulecolor{black}
    \midrule
    \multicolumn{3}{c}{Original passage (relevant)} \\
    \midrule
    \multicolumn{3}{l}{So the built output of Bauhaus architecture in these years is the output of Gropius: the Sommerfeld house}\\
    \multicolumn{3}{l}{in Berlin, the Otte house in Berlin, the Auerbach house in Jena, and the competition design for the Chica-}\\ 
    \multicolumn{3}{l}{go Tribune Tower, which brought the school much attention.taatliches Bauhaus, commonly known simply}\\
    \multicolumn{3}{l}{as Bauhaus, was an art school in Germany that combined crafts and the fine arts, and was famous for the}\\
    \multicolumn{3}{l}{approach to design that it publicised and taught. It operated from 1919 to 1933.}\\
    \midrule
    \multicolumn{3}{c}{Reconstructed relevant (expanded) passage terms }\\ 
    \hline
    \multicolumn{3}{l}{\textcolor{bleudefrance}{ba}, \textcolor{bleudefrance}{\#\#uh}, \textcolor{bleudefrance}{\#\#aus}, ta, output, \textcolor{bleudefrance}{house}, \textcolor{pink}{was}, berlin, year, 1933, \textcolor{bleudefrance}{school}, help, \#\#liche, germany, tribune, architect-}\\
    \multicolumn{3}{l}{ure, jena, chicago, \#\#at, design, 1919, \#\#ius, tower, competition, ot, somme, known, \#\#op, info, art, \#\#bach}\\
    \multicolumn{3}{l}{attention, an, \#\#s, \#\#rf, \textcolor{pink}{location}, commonly, \#\#a, simply, is, for, au, combined, \#\#te, \textcolor{bleudefrance}{built}, operated, \#\#er, as,}\\
    \multicolumn{3}{l}{famous, \#\#ised, brought, and, approach, period, public, \textcolor{pink}{build}, in, fine, \textcolor{pink}{building}, be, reason, crafts, taught, b-}\\
    \multicolumn{3}{l}{ring, greatest, much, were, that, combine, skyscraper, architectural, fact, most, production, largest, are, orig-}\\
    \multicolumn{3}{l}{in, definition, arts, term, so, during, german, name, institution, called, abbreviation, it, type, \#\#t, which, hig-}\\
    \multicolumn{3}{l}{hest, widely, history, of, important, work, operate, war, the, include, state, difference, organization, its, desi-}\\
    \multicolumn{3}{l}{gned, \#\#st, company, great, century, 1920, recent, purpose, sculpture, acronym, meaning, \textcolor{pink}{studio}, produced, }\\
    \multicolumn{3}{l}{last, lot, notable, run, literally, concept, 1934, \textcolor{pink}{college}, \textcolor{pink}{place}, created, record, historical, 1918, clock, major, s-}\\
    \multicolumn{3}{l}{ee, time, decorative, produce, say, being, biggest, \textcolor{pink}{villa}, fifty, many, \textcolor{pink}{founded}, \textcolor{pink}{structure}, culture \ldots (omit)}\\
	\bottomrule
	\end{tabular}
	}

\vspace{0.25cm}

\parbox{\textwidth}{{\small The matching terms between the queries and the original (expanded) passage are colored with blue (red).  The ``rank'' column denotes the rank of the relevant passage under different $\theta$ settings.
	The reconstructed query and passage terms are ordered by their term weights in descending order.}}
\end{table*}

In our final analysis, we demonstrate the interpretability of DLRs, which contrasts with dense semantic representations, where it is often difficult to understand why certain query--passage pairs obtain high scores. 
Since the ``index'' vector stores the position of the most important term in each slice, a DLR can be ``reverted'' back into a bag of words with term weights. 
We showcase how $\theta$ impacts approximate GIP in Table~\ref{tb:showcase}, where we reconstruct expanded query terms from $\deladedlr(768)$, shown in the top portion of the table.
In addition, we reconstruct the passage judged as relevant to the query in the bottom portion of the table.
For simplicity, we do not show term weights.

We observe that when $\theta=0$, there are more than 10 matching terms (colored terms) between the query and the passage. 
As $\theta$ increases, terms with lower weights are filtered out; see Eq.~(\ref{eq:approximate_retrieval}).
Thus, the relevance score between the query and the passage decreases.
For example, from $\theta=0$ to $0.05$, terms with weights lower than $0.05$ are removed for retrieval and the rank of the passage degrades slightly. 
At $\theta=0.3$, four important terms (e.g., `build', `building', `were', `school')\ are removed, and the passage is retrieved much lower in the ranked list (at rank to 19).\footnote{We notice some ``wacky'' terms from the reconstructed passage with high term weights, originally observed by \citet{Mackenzie_etal_arXiv2021}.} 
This example illustrates that tuning $\theta$ determines how many query terms (dimensions) are used for approximate GIP. 

\section{Conclusions and Future Work}

We present a simple yet effective approach to densifying lexical representations for passage retrieval. 
This work introduces a dense representation framework and proposes a new scoring function to compute relevance scores between dense lexical representations (DLRs) derived from queries and passages. 
Using our framework, we can combine lexical and semantic representations into dense hybrid representations (DHRs) for hybrid retrieval. 
Our experiments show that DLRs can accurately approximate any ``off-the-shelf'' lexical model.
Furthermore, when combined with other semantic representations (as DHRs), the resulting models can achieve comparable effectiveness to existing state-of-the-art hybrid retrieval methods.

The main advantage of our framework is that we can execute end-to-end retrieval using DLRs/DHRs on GPUs with a single index structure in a uniform execution environment.
In our implementation, retrieval latency is insensitive to the sparsity of the vectors, unlike lexical representations that use inverted indexes.
This feature makes our approach both fast and easy to deploy, especially for modern lexical retrieval models, which generate representations that are quite dense and require additional tuning (e.g., by introducing a sparsity constraint) to enable practical deployment using inverted indexes. 
Furthermore, we propose to better model content using a single model by jointly training semantic and lexical representations, which are then combined into hybrid representations for retrieval. 
We demonstrate that this combination, $\deladedhr$, outperforms most models (e.g., ANCE, ColBERT, COIL, etc.)\ in both in-domain and out-of-domain evaluations while requiring smaller indexes and achieving lower query latency.
Finally, we examine our two-stage retrieval approach, uncovering how the proposed design achieves low query latency without sacrificing accuracy. 

One future research direction is to combine various supervised techniques~\cite{bpr, jpq, 
RepCONC} for further vector compression. 
In addition, since our proposed single model fusion approach can combine both lexical and semantic matching capabilities, it would be interesting to further explore methods that can integrate the strengths of each in a complementary manner.
Finally, we believe that joint {\it self} training of semantic and lexical components is a promising future direction, especially in domain transfer scenarios.

\section*{Acknowledgements} 

This research was supported in part by the Canada First Research Excellence Fund and the Natural Sciences and Engineering Research Council (NSERC) of Canada.
We acknowledge Cloud TPU support from Google's TPU Research Cloud (TRC).
We thank the anonymous referees who provided useful feedback to improve this work.

\bibliographystyle{ACM-Reference-Format}
\bibliography{paper.bib}

\end{document}

%% file: dlr_comparison.tex
\begin{table}[t]
	\caption{Effectiveness/Efficiency Comparisons of DLRs with Different Base Models on MS MARCO (Dev)}
	\label{tb:dim_reduction}
	\centering
	\resizebox{0.8\width}{!}{  
	\begin{threeparttable}
    \begin{tabular}{ll!{\color{lightgray}\vrule}c!{\color{lightgray}\vrule}l!{\color{lightgray}\vrule}l!{\color{lightgray}\vrule}llrr}
	\toprule
& \multirow{2}{*}{Base model} &\multirow{2}{*}{Method}& \multirow{2}{*}{Dim} & \ \ \# tokens/doc & \multicolumn{2}{c}{Quality} &Storage & Latency \\ 
 \cmidrule(lr){5-5} \cmidrule(lr){6-7} \cmidrule(lr){8-8} \cmidrule(lr){9-9} 
 &	& &  & number (diff.) & MRR@10 (diff.)& R@1K  (diff.)& GB & ms/q \\
\midrule
 \multirow{9}{*}{\rotatebox[origin=c]{90}{whole word}} &\multirow{5}{*}{BM25} 
 & Inv. index& 2.6M  & \ 30.11 & 0.188& 0.858& 0.7 & 40 \\
 	 \arrayrulecolor{lightgray}
 \cline{3-9}
 & & SPAR~\cite{spar} & 768 & \ -  & 0.173 \ \ ($-$8.0\%)  & 0.831 \ \ ($-$3.1\%)& 26.0 & 64\\ 
  \cline{3-9}
   & & \multirow{3}{*}{DLR}  & 768  & \ 29.18 \ \ ($-$3.1\%)  & 0.180 \ \ ($-$4.3\%) & 0.845 \ \ ($-$1.5\%)& 26.0 & 26\\
   &   &  &  256 & \ 28.41 \ \ ($-$5.7\%) & 0.177 \ \ ($-$5.9\%)& 0.834 \ \ ($-$2.8\%)& 8.6 & 23\\
   &   & & 128 & \ 26.62 ($-$11.6\%) & 0.169 ($-$10.1\%) & 0.816 \ \ ($-$4.9\%)& 4.3 & 22\\ 
      \arrayrulecolor{black}
      \cline{2-9}
 &\multirow{4}{*}{DeepImpact} 
 & Inv. index& 3.5M  &  \ 71.61  & 0.327 & 0.948 & 1.4 & 285\\
 	 \arrayrulecolor{lightgray}
 \cline{3-9}
  &  & \multirow{3}{*}{DLR}  & 768 & \ 65.28 \ \ ($-$8.9\%) & 0.324 \ \ ($-$0.9\%) & 0.942 \ \ ($-$0.6\%) & 26.0 & 26\\
    &  &  & 256 & \ 59.90\ ($-$16.4\%) & 0.316 \ \ ($-$3.4\%) & 0.933 \ \ ($-$1.6\%) & 8.6 & 24\\
  &    &  & 128 & \ 52.74 ($-$26.4\%)  & 0.304 \ \ ($-$7.0\%) & 0.923 \ \ ($-$2.6\%) & 4.3 & 22\\ 
      \arrayrulecolor{black}
      \midrule 
 \multirow{13}{*}{\rotatebox[origin=c]{90}{wordpiece token}} &\multirow{5}{*}{uniCOIL} 
  & Inv. index& 30K  & \ 67.96  & 0.351 & 0.958 & 1.3 & 291\\
 	 \arrayrulecolor{lightgray}
 \cline{3-9}
 &  & SPAR~\cite{spar} & 768 & \ -  & 0.341 \ \ ($-$2.8\%)  & 0.970 \ \ ($+$1.2\%) & 26.0 & 64\\ 
  \cline{3-9}
 &   & \multirow{3}{*}{DLR}  & 768  & \ 64.15 \ \ ($-$5.6\%)  & 0.349 \ \ ($-$0.6\%) & 0.957 \ \ ($-$0.1\%)  & 20.0 & 25 \\
   &   & &  256 & \ 58.62 ($-$13.7\%) & 0.344 \ \ ($-$2.0\%)  & 0.952 \ \ ($-$0.6\%)  & 6.4 & 22\\
    &  & & 128  & \ 52.48 ($-$22.8\%) & 0.335 \ \ ($-$4.6\%)  & 0.944 \ \ ($-$1.5\%)  & 3.3 & 22 \\
      \arrayrulecolor{black}
      \cline{2-9}
     &   \multirow{4}{*}{SPLADE} & Inv. index & 30K & \ 91.50  & 0.340 & 0.965 & 2.6 & 475 \\  
    \arrayrulecolor{lightgray}
    \cline{3-9}
   &   & \multirow{3}{*}{DLR}& 768  & \ 86.33 \ \ ($-$5.7\%)  & 0.336 \ \ ($-$1.2\%) & 0.963 \ \ ($-$0.2\%) & 20.0 & 28\\
   &   & & 256 & \ 76.45 ($-$16.5\%) & 0.326 \ \ ($-$4.2\%) & 0.959 \ \ ($-$0.6\%)  & 6.4 & 25\\
   &   &  & 128 & \ 64.35 ($-$29.7\%) & 0.318 \ \ ($-$6.5\%) & 0.951 \ \ ($-$1.5\%)  & 3.3 & 24\\
       \arrayrulecolor{black}
     \cline{2-9}
  &  \multirow{4}{*}{DeLADE}  & \texttt{FlatIP}\tnote{$\star$}  & 30K  & 30522   & 0.347& 0.957& 1033.3& - \\
    \arrayrulecolor{lightgray}
    \cline{3-9}
   &  & \multirow{3}{*}{DLR} & 768& \ \ \ \ 768 ($-$97.5\%)& 0.345 \ \ ($-$0.6\%) & 0.953 \ \ ($-$0.4\%) & 20.0& 25\\
  &     & & 256 & \ \ \ \ 256 ($-$99.2\%)  & 0.341 \ \ ($-$1.7\%) & 0.951 \ \ ($-$0.6\%)  & 6.4 & 22  \\
   &    &  & 128 & \ \ \ \ 128 ($-$99.6\%) & 0.335 \ \ ($-$3.5\%) & 0.945 \ \ ($-$1.3\%)  & 3.3 & 21\\
         \arrayrulecolor{black}
	\bottomrule
	\end{tabular}
	\begin{tablenotes}
	    \item[$\star$] Index size with \texttt{faiss.FlatIP} is provided only as a reference; query latency is not comparable to retrieval with inverted indexes and hence omitted.
    \end{tablenotes}
    \end{threeparttable}}
\end{table}

%% file: compression_comparison.tex
\begin{table}[t]
	\caption{Effectiveness Comparisons of DLRs with Different Unsupervised Compression Techniques on FiQA-2018 (Test).}
	\label{tb:unsupervised_dim_reduction}
	\centering
	\resizebox{0.8\width}{!}{
    \begin{tabular}{lllccc}
	\toprule
\multirow{2}{*}{Base model} &\multirow{2}{*}{Method}& \multirow{2}{*}{Dim} &  \multicolumn{2}{c}{Qualtiy}& Storage \\ 
\cmidrule(lr){4-5} \cmidrule(lr){6-6} 
 & &  &nDCG@10 (diff.) &R@100 (diff.)& GB \\
\midrule
 \multirow{7}{*}{DeLADE}  & \texttt{FlatIP}  & 30K& 0.301 \ \ \ \ \ \ \ \ \ \ \ \ \ \ \ & 0.576 \ \ \ \ \ \ \ \ \ \ \ \ \ \ \ & 6.54 \\
    \arrayrulecolor{lightgray}
    \cline{2-6}
& \texttt{LSH}& -& 0.210 ($-$30.2\%)& 0.545 \ \ ($-$5.4\%)& 0.75\\
 \cline{2-6}
& \multicolumn{1}{l}{\texttt{PQ768}}& -& 0.281 \ \ ($-$6.6\%)& 0.557 \ \ ($-$3.3\%)& 0.07\\
& \multicolumn{1}{l}{\texttt{PQ256}}& -& 0.253 ($-$15.9\%)& 0.535 \ \ ($-$7.1\%)& 0.04\\
& \multicolumn{1}{l}{\texttt{PQ128}}& -& 0.221 ($-$26.6\%)& 0.496 ($-$13.9\%)& 0.03\\
 \cline{2-6}
& \multirow{3}{*}{DLR} & 768&0.294 \ \ ($-$2.3\%)& 0.567 \ \ ($-$1.6\%)& 0.13\\
& & 256& 0.287 \ \ ($-$4.7\%)& 0.558 \ \ ($-$3.1\%)& 0.04\\
&  & 128& 0.274 \ \ ($-$9.0\%)& 0.552 \ \ ($-$4.2\%)& 0.02\\
         \arrayrulecolor{black}
	\bottomrule
	\end{tabular}}
\end{table}

%% file: dhr_comparison.tex
\begin{table*}[t]
	\caption{Effectiveness/Efficiency Comparisons of Independent Fusion of DLRs with ``Off-the-Shelf'' Dense Semantic Representations on MS MARCO (Dev)}
	\label{tb:fusion_retrieval_comparison}
	\centering
	\resizebox{1\textwidth}{!}{  
    \begin{tabular}{lcccr!{\color{lightgray}\vrule}ll!{\color{lightgray}\vrule}cc}
	\toprule
    \multirow{2}{*}{Approach} & \multicolumn{2}{c}{Lexical} & \multicolumn{2}{c}{Semantic} &\multicolumn{2}{c}{qualtiy} & \multicolumn{1}{c}{storage} & \multicolumn{1}{c}{latency}   \\
\cmidrule(lr){2-3}\cmidrule(lr){4-5} \cmidrule(lr){6-7} \cmidrule(lr){8-8} \cmidrule(lr){9-9}
 &	Software & Dim & Software & Dim  & MRR@10 (diff.)& R@1K (diff.)& (GB) & (ms/q) \\
\midrule 
{\small(1)}~Linear combination (BM25, ANCE) & Lucene & 30K & Faiss \texttt{FlatIP} & 768 & 0.347 & 0.969 & 26 & \ \ 64 \\ 
\arrayrulecolor{lightgray}
   \midrule
 {\small(2)}~SPAR~\cite{spar} (lexical conversion to semantic) & n/a & n/a & Faiss \texttt{FlatIP} & $2\times$ 768  & 0.344 \ ($-$0.9\%)  & 0.971 \ ($+$0.2\%) & 52 & \ \ 81\\
 \midrule
  & PyTorch & 768 &PyTorch  & 768  & 0.349 \ ($-$0.6\%) & 0.967 \ ($-$0.2\%) & 39 &  \ \ 56 \\
 {\small(3)}~$(\text{BM25+ANCE})_{\text{\tiny DHR}}$ & PyTorch & 256  &PyTorch & 768 & 0.348 \ ($+$0.3\%) & 0.967 \ ($-$0.2\%) & 21 &  \ \ 56 \\
    & PyTorch  & 128& PyTorch& 768 & 0.347 \ ($-$0.0\%)  & 0.967 \ ($-$0.2\%)  & 17 &  \ \ 53 \\ 
 \arrayrulecolor{black}
\midrule
 {\small(4)}~Linear combination (uniCOIL, ANCE) & Lucene & 30K & Faiss \texttt{FlatIP} & 768 & 0.375 & 0.976 & 27 & 291 \\ 
\arrayrulecolor{lightgray}
\midrule
 {\small(5)}~SPAR~\cite{spar} (lexical conversion to semantic) & n/a & n/a & Faiss \texttt{FlatIP} & $2\times$ 768  & 0.369 \ ($-$1.6\%) & 0.981 \ ($+$0.5\%) & 52 & \ \ 81\\
 \midrule
  & PyTorch & 768 &PyTorch  & 768  & 0.378 \ ($+$0.8\%) & 0.975 \ ($-$0.1\%) & 32 &  \ \ 60 \\
 {\small(6)}~$(\text{uniCOIL+ANCE})_{\text{\tiny DHR}}$ & PyTorch & 256  &PyTorch & 768 & 0.375 \ ($-$0.0\%) & 0.973 \ ($-$0.3\%) & 19 &  \ \ 58 \\
    & PyTorch  & 128& PyTorch& 768& 0.369 \ ($-$1.6\%) & 0.971 \ ($-$0.5\%) & 16 &  \ \ 57 \\ 
     \arrayrulecolor{black}
	\bottomrule
	\end{tabular}
    }

\vspace{0.25cm}

\parbox{\textwidth}{{\small Retrieval with DHRs uses our custom PyTorch implementation running on GPUs.}}
 
\end{table*}

%% file: main_comparison.tex
\begin{table*}[t]
\caption{Effectiveness/Efficiency Comparisons of DHRs Using Single Model Fusion 
	}
	\label{tb:main_result}
	\centering
	 \resizebox{1\textwidth}{!}{  
	 \begin{threeparttable}
	 \setlength\tabcolsep{3.65pt}
    \begin{tabular}{llllllllllllll}
	\toprule
 &  & \multicolumn{3}{c}{sparse lexical} & \multicolumn{2}{c}{dense semantic} & \multicolumn{2}{c}{multi-vector}  & $\deladedlr$ & \multicolumn{3}{c}{$\deladedhr$}\\
 \cmidrule(lr){3-5} \cmidrule(lr){6-7}\cmidrule(lr){8-9} \cmidrule(lr){10-10} \cmidrule(lr){11-13}
& & {\small(1)}~BM25& {\small(2)}~docT5q & {\small(3)}~SPLADE&  {\small(4)}~Dense&  {\small(5)}~ANCE&  {\small(6)}~COIL &  {\small(7)}~ColBERT&  {\small(8)}~768 dim&  {\small(9)}~128 dim&  {\small(a)}~256 dim&  {\small(b)}~768 dim  \\
 \midrule
\multicolumn{2}{l}{\textbf{Efficiency}\tnote{$\ast$}} \\
\arrayrulecolor{lightgray}
\midrule 
\multicolumn{2}{l}{storage (GB)} & 0.67 & 0.98 & 2.6 & 26 & 26  & 60 & 154  & 20  & 5.4  & 8.6  & 22 \\
\multicolumn{2}{l}{latency (ms/q)} & 40 & 64  & 475  & 64  &  64  & 40\tnote{$\star$} & 458\tnote{$\star$} & 30  & 28  &  31  & 33 \\
\arrayrulecolor{black}
\midrule 
\multicolumn{2}{l}{\textbf{MS MARCO}} \\
\arrayrulecolor{lightgray}
\midrule 
\multirow{2}{*}{Dev}& \small MRR@10 & 0.188 & 0.277$^{\tiny1}$ & 0.340$^{\tiny1245}$ & 0.307$^{\tiny12}$ & 0.330$^{\tiny124}$ & 0.354$^{\tiny1-58}$ & \textbf{0.360}\tnote{$\bigtriangleup$} & 0.345$^{\tiny1245}$ & 0.351$^{\tiny1-58}$  & 0.355$^{\tiny1-589}$ & \underline{0.357}$^{\tiny1-58-\text{a}}$ \\
& \small R@1K & 0.858 & 0.947$^{\tiny1}$ & 0.965$^{\tiny12458}$ & 0.944$^{\tiny1}$ & 0.959$^{\tiny124}$ & 0.964$^{\tiny12458}$ & \textbf{0.968}\tnote{$\bigtriangleup$} & 0.953$^{\tiny124}$& 0.962$^{\tiny1248}$ & 0.965$^{\tiny124589}$ & \underline{0.967}$^{\tiny12458-\text{a}}$\\
\midrule 
DL 19 & \multirow{2}{*}{\small nDCG@10} & 0.506 & 0.642$^{\tiny1}$ & 0.683$^{\tiny14}$ & 0.631$^{\tiny1}$ & 0.646$^{\tiny1}$& \textbf{0.714}$^{\tiny1245}$ & 0.694\tnote{$\bigtriangleup$}& 0.691$^{\tiny14}$ & 0.691$^{\tiny145}$  & \underline{0.696}$^{\tiny145}$ & 0.693$^{\tiny145}$\\
DL 20& & 0.475 & 0.619$^{\tiny1}$ & 0.671$^{\tiny1}$ & 0.648$^{\tiny1}$ & 0.646$^{\tiny1}$ & \textbf{0.688}$^{\tiny125}$& 0.676\tnote{$\bigtriangleup$} & 0.668$^{\tiny12}$& 0.683$^{\tiny12}$ & \underline{0.686}$^{\tiny12}$ & \textbf{0.688}$^{\tiny12}$\\
\arrayrulecolor{black}
\midrule  
\multicolumn{2}{l}{\textbf{BEIR}}& \multicolumn{12}{c}{nDCG@10}\\
\arrayrulecolor{lightgray}
\midrule  
\multicolumn{2}{l}{TREC-COVID}   & 0.656& \underline{0.713}&	0.661& 0.604& 0.654& 0.668&	0.677&	0.681&	0.695&	0.701 & \textbf{0.727}\\
\multicolumn{2}{l}{NFCorpus}& 0.325&	0.328&	0.322& 0.244& 0.237&	\textbf{0.331}& 0.305&	\underline{0.331}&	0.319&	0.324& 0.329\\
\multicolumn{2}{l}{NQ}& 0.329&	0.399&	0.469& 0.410&	0.446& \underline{0.519}&	\textbf{0.524}&	0.471&	0.476&	0.487 & 0.497\\
\multicolumn{2}{l}{HotpotQA}& 0.603&	0.580&	0.640& 0.441&	0.456& \textbf{0.713}&	0.593&	0.666&	0.664&	0.679 & \underline{0.689}\\
\multicolumn{2}{l}{FiQA-2018}& 0.236&	0.291&	0.289& 0.224& 0.295& \underline{0.313}&	\textbf{0.317}&	0.294&	0.292&	0.304& 0.312\\
\multicolumn{2}{l}{ArguAna}&0.315&	0.349&	\textbf{0.445}& 0.323&	0.415& 0.295&	0.233&	0.360&	\underline{0.423}&	0.405 & 0.360\\
\multicolumn{2}{l}{Tóuche-2020 {\small(v2)}}& \textbf{0.367}&	\underline{0.347}&	0.201& 0.185&	0.240& 0.281&	0.202&	0.266&	0.248&	0.259& 0.280\\
\multicolumn{2}{l}{Quora}& 0.789&	0.802&	0.834& 0.750&	\underline{0.852}& 0.838&	\textbf{0.854}&	0.755&	0.829&	0.830& 0.831\\
\multicolumn{2}{l}{DBPedia}& 0.313&	0.331&	0.370& 0.295& 0.281& 0.398&	0.392&	0.376&	0.397&	\underline{0.402} & \textbf{0.404}\\
\multicolumn{2}{l}{Scidocs}&0.158&	\underline{0.162}&	0.149& 0.103&	0.122&	0.155& \textbf{0.165}&	0.148&	0.146&	0.148 & 0.150\\
\multicolumn{2}{l}{Fever}& 0.753&	0.714&	0.740& 0.651&	0.669&	\textbf{0.840}& 0.771&	0.787&	0.750&	0.781 & \underline{0.810}\\
\multicolumn{2}{l}{Climate-Fever}& 0.213&	0.201&	0.187& 0.167&	0.198& 0.216& 0.184&	0.202&	0.215&	\underline{0.220} & \textbf{0.229}\\
\multicolumn{2}{l}{SciFact}& 0.665&	0.675&	0.633& 0.479&	0.507& \textbf{0.707}&	0.671&	0.674&	0.651&	0.670 & \underline{0.685}\\
\arrayrulecolor{lightgray}
\midrule 
\multicolumn{2}{l}{Avg. nDCG@10}&0.440&	0.453&	0.458& 0.375&	0.413& \underline{0.483}&	0.453&	0.462&	0.470&	0.478 & \textbf{0.485}\\
\multicolumn{2}{l}{Avg. rank}&7.00	&6.07	&7.00	&10.46	&8.15	&\underline{3.15}	&5.54	&5.39	&5.92	&4.46	&\textbf{2.85} \\
\arrayrulecolor{black}
\midrule  
\multicolumn{2}{l}{\textbf{BEIR}}& \multicolumn{12}{c}{R@100}\\
\arrayrulecolor{lightgray}
\midrule  
\multicolumn{2}{l}{TREC-COVID} &	0.498&	\textbf{0.541}&	0.502&	0.386&	0.457& 0.531&	0.464&	0.502&	0.489&	\underline{0.512}& \textbf{0.541}\\
\multicolumn{2}{l}{NFCorpus} &	0.250&	0.253&	0.266&	0.228&	0.232&	\textbf{0.277}&	0.254&	0.260&	0.268&	0.272& \underline{0.273} \\
\multicolumn{2}{l}{NQ} &	0.760&	0.832&	0.883&	0.817&	0.836& \textbf{0.917}&	0.912&	0.876&	0.894&	0.901& \underline{0.909} \\
\multicolumn{2}{l}{HotpotQA} &	0.740&	0.709&	0.793&	0.586&	0.578& \textbf{0.837}&	0.748&	0.801&	0.797&	0.808& \underline{0.818} \\
\multicolumn{2}{l}{FiQA-2018}&	0.539&	0.598&	0.576&	0.500&	0.581&	\underline{0.596}&	\textbf{0.603}&	0.567&	0.584&	0.592& 0.591 \\
\multicolumn{2}{l}{ArguAna} &	0.942&	\textbf{0.972}&	0.954&	0.873&	0.937&	0.777&	0.914&	0.893&	\underline{0.957}&	0.943& 0.900 \\
\multicolumn{2}{l}{Touché-2020 (v2)} &	\underline{0.538}&	\textbf{0.557}&	0.450&	0.409&	0.458& 0.512&	0.439&	0.499&	0.462&	0.480& 0.500 \\
\multicolumn{2}{l}{Quora} &	0.973&	0.982&	0.984&	0.963&	0.987& \textbf{0.998}&	\underline{0.989}&	0.972&	0.984&	0.984& 0.985 \\
\multicolumn{2}{l}{DBPedia} &	0.398&	0.365&	0.493&	0.353&	0.319& \textbf{0.522}&	0.461&	0.495&	0.491&	0.506& \underline{0.510} \\
\multicolumn{2}{l}{SCIDOCS} &	\underline{0.356}&	\textbf{0.360}&	0.351&	0.247&	0.269& 0.354&	0.344&	0.343&	0.338&	0.342& 0.346 \\
\multicolumn{2}{l}{FEVER} &	0.931&	0.916&	0.934&	0.886&	0.900& \textbf{0.959}&	0.934&	0.948&	0.948&	0.953& \underline{0.956} \\
\multicolumn{2}{l}{Climate-FEVER} &	0.436&	0.427&	0.452&	0.401&	0.445& \underline{0.508}&	0.444&	0.464&	0.490&	0.506& \textbf{0.525} \\
\multicolumn{2}{l}{SciFact} &	0.908&	0.914&	0.899&	0.829&	0.816& \textbf{0.931}&	0.878&	\underline{0.920}&	0.898&	0.904& 0.910 \\
\arrayrulecolor{lightgray}
\midrule 
\multicolumn{2}{l}{Avg. R@100} &	0.636&	0.648& 0.657&	0.575&	0.601& \underline{0.671}&	0.645&	0.657&	0.661&	0.669& \textbf{0.674} \\
\multicolumn{2}{l}{Avg. rank}&7.15&	5.54&	6.08&	10.62	&8.62	&\textbf{2.46}	&6.23	&5.92	&5.77	&4.31	&\underline{3.31} \\
\arrayrulecolor{black}
	\bottomrule
	\end{tabular}
	\begin{tablenotes}
	\item[$\ast$] We report efficiency figures on the MS MARCO passage corpus for comparison.
	\item[$\star$] These numbers are copied from the original papers, which are measured on multi-GPU systems; thus, they cannot be reproduced in our setup.
	\item[$\bigtriangleup$] We do not have these run files; thus, no significance testing is performed against ColBERT.
    \end{tablenotes}
    \end{threeparttable}
	}

\vspace{0.25cm}

\parbox{\textwidth}{{\small This table compares against a selection of retrieval models trained with comparable baseline strategies.
Bold (underline) denotes the best (second best) effectiveness for each row. 
For the MS MARCO datasets, superscripts denote significant improvements over the labeled model with paired $t$-test ($p <0.05$).}}
 
\end{table*}

\begin{table}[t]
	\caption{Component Retrieval Effectiveness of $\deladedhr(768)$}
	\label{tb:joint_training_effect}
	\centering
	\resizebox{0.8\width}{!}{  
	\begin{threeparttable}
    \begin{tabular}{lclllllll}
	\toprule
 &\multicolumn{2}{c}{MS MARCO dev}&\multicolumn{2}{c}{TREC-COVID} &\multicolumn{2}{c}{FiQA-2018} &\multicolumn{2}{c}{SciFact} \\
	\cmidrule(lr){2-3} \cmidrule(lr){4-5} \cmidrule(lr){6-7}\cmidrule(lr){8-9} 
	 Component\tnote{$\star$}&  MRR@10&  R@1K   &  NDCG@10&  CapR@100  &  NDCG@10&  R@100  & NDCG@10&  R@100 \\
\midrule
 {\small(1)}~DeLADE + \texttt{[CLS]}& 0.357& 0.967& 0.727& 0.541& 0.312& 0.591& 0.685& 0.910\\ 
 \arrayrulecolor{lightgray}
 \hline
 \arrayrulecolor{black}
 {\small(2)}~DeLADE& 0.294& 0.930& 0.658& 0.501& 0.193& 0.441& 0.684& 0.920 \\ 
 {\small(3)}~\texttt{[CLS]}& 0.045& 0.494& 0.125& 0.091& 0.037& 0.185& 0.118& 0.454\\
	\bottomrule
	\end{tabular}
	\begin{tablenotes}
	\item[$\star$] Row (1) corresponds to $\deladedhr(768)$ in Table~\ref{tb:main_result}. 
	Note that rows (2) and (3) are different from the Dense and $\deladedhr(768)$ conditions in Table~\ref{tb:main_result}, where the separate models are trained independently.
    \end{tablenotes}
    \end{threeparttable}
	}
\end{table}

\begin{table}[ht]
	\caption{Ablation of $\deladedhr(768)$ Varying the Semantic Component}
	\label{tb:cls_ablation}
	\centering
	\resizebox{0.8\width}{!}{  
	\begin{threeparttable}
    \begin{tabular}{lllllllll}
	\toprule
 &\multicolumn{2}{c}{MS MARCO dev}&\multicolumn{2}{c}{TREC-COVID} &\multicolumn{2}{c}{FiQA-2018} &\multicolumn{2}{c}{SciFact} \\
	\cmidrule(lr){2-3} \cmidrule(lr){4-5} \cmidrule(lr){6-7}\cmidrule(lr){8-9} 
	$\text{[CLS]}$ dimension&  MRR@10&  R@1K   &  NDCG@10&  CapR@100  &  NDCG@10&  R@100  & NDCG@10&  R@100 \\
\midrule
        {\small(1)}~0\tnote{$\star$} & 0.345 & 0.953& 0.681 & 0.502 & 0.294 & 0.567& 0.674& \textbf{0.920}\\
        {\small(2)}~128\tnote{$\star$} & 0.357$^{\tiny 1}$& 0.967$^{\tiny 1}$& \textbf{0.727}$^{\tiny 1}$ & \textbf{0.541}$^{\tiny 1}$ & 0.312$^{\tiny 1}$ &  0.591$^{\tiny 1}$& \textbf{0.685} & 0.910\\
        {\small(3)}~256 & 0.358$^{\tiny 1}$ & 0.969$^{\tiny 1}$& 0.722$^{\tiny 1}$ & 0.530$^{\tiny 1}$ & 0.313$^{\tiny 1}$ & 0.590$^{\tiny 1}$ & 0.683 & 0.917\\
        {\small(4)}~768 & 0.358$^{\tiny 1}$& 0.969$^{\tiny 1}$& 0.717 & 0.540$^{\tiny 1}$ & \textbf{0.318}$^{\tiny 1}$ & \textbf{0.601}$^{\tiny 1}$ & 0.680 & 0.918\\
	\bottomrule
	\end{tabular}
	\begin{tablenotes}
	\item[$\star$] The 0 and 128 variants correspond to $\deladedlr(768)$ and $\deladedhr(768)$ in Table~\ref{tb:main_result}, respectively.
    \end{tablenotes}
    \end{threeparttable}
	}

\vspace{0.25cm}

\parbox{\textwidth}{{\small Bold denotes the best effectiveness for each column. Superscript denotes significant improvement over the labeled model based on paired $t$-tests ($p<0.05$).}}
\end{table}

%% file: advanced_comparison.tex
\begin{table*}[t]
	\caption{Effectiveness/Efficiency Comparisons with Existing State-of-the-Art Models Using Advanced Training Techniques}
	\label{tb:advanced_comparison}
	\centering
	 \resizebox{1\textwidth}{!}{  
	 \begin{threeparttable}
	  	 \setlength\tabcolsep{2.5pt}
    \begin{tabular}{llccccccccccc}
	\toprule
\multicolumn{2}{c}{}&
\multicolumn{1}{c}{sparse lexical}&
\multicolumn{6}{c}{dense semantic} &  \multicolumn{1}{c}{multi-vector} & \multicolumn{3}{c}{$\deladedhr^+$}\\
\cmidrule(lr){3-3}
 \cmidrule(lr){4-9}
 \cmidrule(lr){10-10}
 \cmidrule(lr){11-13}
 \multicolumn{2}{l}{}&
 \multicolumn{1}{c}{{\small(c)}~SPLADEv2}&
 \multicolumn{1}{c}{{\small(d)}~TAS-B}& \multicolumn{1}{c}{{\small(e)}~Contriever}& \multicolumn{1}{c}{{\small(f)}~RocketQAv2}& \multicolumn{1}{c}{{\small(g)}~GPL} & \multicolumn{1}{c}{{\small(h)}~GTR\tiny XL}& \multicolumn{1}{c}{{\small(i)}~GTR\tiny XXL}& \multicolumn{1}{c}{{\small(j)}~ColBERTv2} & {\small(k)}~128 dim& {\small(l)}~256 dim& {\small(m)}~768 dim  \\
 \midrule
  \arrayrulecolor{lightgray}
 \multicolumn{2}{l}{model size} & 66M &  66M & 110M & 110M& 66M$\times 13$\tnote{$\star$}& 1.24B & 4.8B & 110M & & 66M &\\
 \midrule 
  \multicolumn{2}{l}{IR pretrain} &\xmark  & \xmark &  \cmark& \xmark & \xmark & \cmark & \cmark  &\xmark &  &\xmark & \\
  \midrule 
 \multicolumn{2}{l}{KD}&  \cmark  & \cmark & \xmark &  \cmark &  \cmark &  \cmark& \cmark& \cmark &  &\cmark & \\
 \midrule 
 \multicolumn{2}{l}{HNM}& \cmark  & \xmark & \cmark& \cmark & \cmark & \cmark & \cmark & \cmark&  &\cmark &\\
 \midrule 
 \multicolumn{2}{l}{batch size >1K} & \xmark  & \xmark & \cmark& \cmark & \xmark & \cmark & \cmark  & \xmark&  &\xmark &\\
 \arrayrulecolor{black}
\midrule 
\multicolumn{12}{l}{\textbf{Efficiency}\tnote{$\ast$}}  \\
\arrayrulecolor{lightgray}
\midrule 
\multicolumn{2}{l}{storage (GBs)} & 5.0  & 26 & 26 & 26 & 26 & 26 & 26 & 29& 5.4 & 8.6 & 22\\
\multicolumn{2}{l}{latency (ms/q)}& 2864 & 64 & 64 & 64 & 64 & 64 & 64 & 260& 28 & 31 & 33 \\
\arrayrulecolor{black}
\midrule 
\multicolumn{2}{l}{\textbf{MS MARCO}} \\
\arrayrulecolor{lightgray}
\midrule 
\multirow{2}{*}{Dev}& MRR@10&	0.368& 0.347&	0.341& 0.381& -&	0.385&	\underline{0.388}&	\textbf{0.397}&	0.366& 0.370& 0.371 \\
& R@1K&	\underline{0.979}& 0.978&	0.979& 0.981&	-&	\underline{0.989}&	\textbf{0.990}& 0.984&	0.973&	0.975&	0.977\\
\midrule 
DL 19 & \multirow{2}{*}{nDCG@10}&	\textbf{0.729}& \underline{0.717}&	0.678&	-&	-&	-&	-&	-&	0.703&	0.711&	0.708\\
DL 20& &	\textbf{0.710}& 0.686&	0.661&	-&	-&	-&	-&	-&	0.684&	0.696&	\underline{0.700}\\
\arrayrulecolor{black}
\midrule  
\multicolumn{2}{l}{\textbf{BEIR}} & \multicolumn{11}{c}{nDCG@10}\\
\arrayrulecolor{lightgray}
\midrule  
\multicolumn{2}{l}{TREC-COVID}&	0.710&  0.481&	0.596& 0.675&	0.700&	0.584&	0.501&	\textbf{0.738}&	0.686&	0.702&	\underline{0.735}\\
\multicolumn{2}{l}{NFCorpus}&	0.334& 	0.319&	0.328& 0.293& \textbf{0.345}&	\underline{0.343}&	0.342&	0.338&	0.327&	0.332&	0.337\\
\multicolumn{2}{l}{NQ}&	0.521& 0.463&	0.498& 0.505&	0.483&	0.559&	\textbf{0.568}&	\underline{0.562}&	0.497& 0.512& 0.523\\
\multicolumn{2}{l}{HotpotQA}&	\textbf{0.684}& 	0.584&	0.638& 0.533&	0.582&	0.591&	0.599&	0.667&	0.659&	\underline{0.673}&	\textbf{0.684}\\
\multicolumn{2}{l}{FiQA-2018}&	0.336& 0.300&	0.329&	0.302& 0.344&	\underline{0.444}&	\textbf{0.467}&	0.356&	0.320&	0.326&	0.335\\
\multicolumn{2}{l}{ArguAna}&	0.479& 0.429&	0.446&	0.451& \textbf{0.557}&	0.531&	\underline{0.540}&	0.463&	0.475&	0.458&	0.436\\
\multicolumn{2}{l}{Tóuche-2020 {\small(v2)}}&	\textbf{0.272}& 0.162&	0.230&	0.247& 0.255&	0.230&	0.256&	\underline{0.263}&	0.226&	0.237&	0.254\\
\multicolumn{2}{l}{Quora}&	0.838& 0.835&	0.865&	0.749& 0.836&	\underline{0.890}&	\textbf{0.892}&	0.852&	0.846&	0.848&	0.849\\
\multicolumn{2}{l}{DBPedia}&	\underline{0.435}& 0.384&	0.413&	0.356& 0.384&	0.396&	0.408&	\textbf{0.446}&	0.402&	0.409&	0.413\\
\multicolumn{2}{l}{Scidocs}&	0.158& 0.149&	\underline{0.165}& 0.131&	\textbf{0.169}&	0.159&	0.161&	0.154&	0.156&	0.158&	0.158\\
\multicolumn{2}{l}{Fever}&	0.786& 0.700&	0.758& 0.676&	0.759&	0.717&	0.740&	0.785&	0.764&	\underline{0.794}&	\textbf{0.815}\\
\multicolumn{2}{l}{Climate-Fever}&	0.235& 0.228&	0.237& 0.180& 0.235&	\textbf{0.270}&	\underline{0.267}&	0.176&		0.222& 0.232&	0.239\\
\multicolumn{2}{l}{SciFact}&	0.693& 0.643&	0.677& 0.568&	0.674&	0.635&	0.662&	0.693&	0.674&	\underline{0.694}&	\textbf{0.699}\\
\arrayrulecolor{lightgray}
\midrule
\multicolumn{2}{l}{Avg.nDCG@10}&	\textbf{0.499}& 0.437&	0.475& 0.436&	0.486&	0.488&	0.493&	\textbf{0.499}&	0.481& 0.490&	\underline{0.498}\\
\multicolumn{2}{l}{Avg. rank}&	\underline{4.15}	&10.00	&6.00	&9.62	&5.46 &5.31	&4.23	&\underline{4.15}	&7.31	&5.38	&\textbf{4.00}\\
\arrayrulecolor{black}
	\bottomrule
	\end{tabular}
		\begin{tablenotes}
	\item[$\star$] \citet{gpl} trained one expert model for each BEIR dataset using pseudo-relevant labels from a cross-encoder model.
	\item[$\ast$] We report efficiency figures on the MS MARCO passage corpus for comparison. As detailed in \citet{Mackenzie_etal_arXiv2021}, SPLADEv2 is much slower than the other models with Lucene; even with PISA, a much faster query evaluation implementation, latency is still 220 ms/q. 
	As for ColBERTv2, the number is copied from Vanilla ColBERTv2 (p=4, c=$2^{16}$) measured by \citet{plaid}. 
    \end{tablenotes}
    \end{threeparttable}
	}

\vspace{0.25cm}

 Bold (underline) denotes the best (second best) effectiveness for each row.

\end{table*}

%% file: two_stage_figures.tex
\begin{figure}[t]
\begin{subfigure}[b]{0.49\columnwidth}
\caption{$\deladedlr(768)$}
\includegraphics[width=\columnwidth]{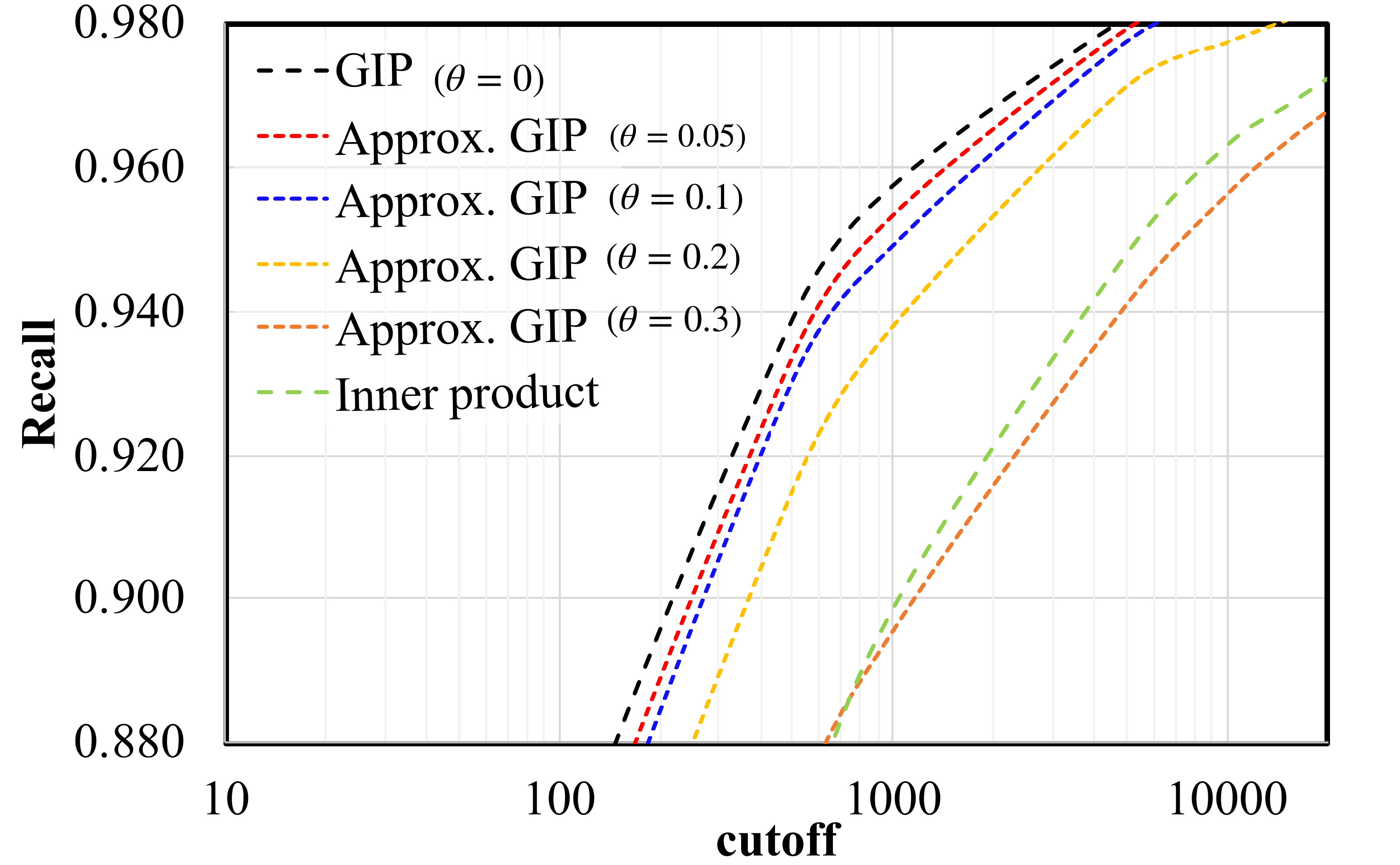}
\end{subfigure}
\begin{subfigure}[b]{0.49\columnwidth}
\caption{$\deladedhr(768)$}
\includegraphics[width=\columnwidth]{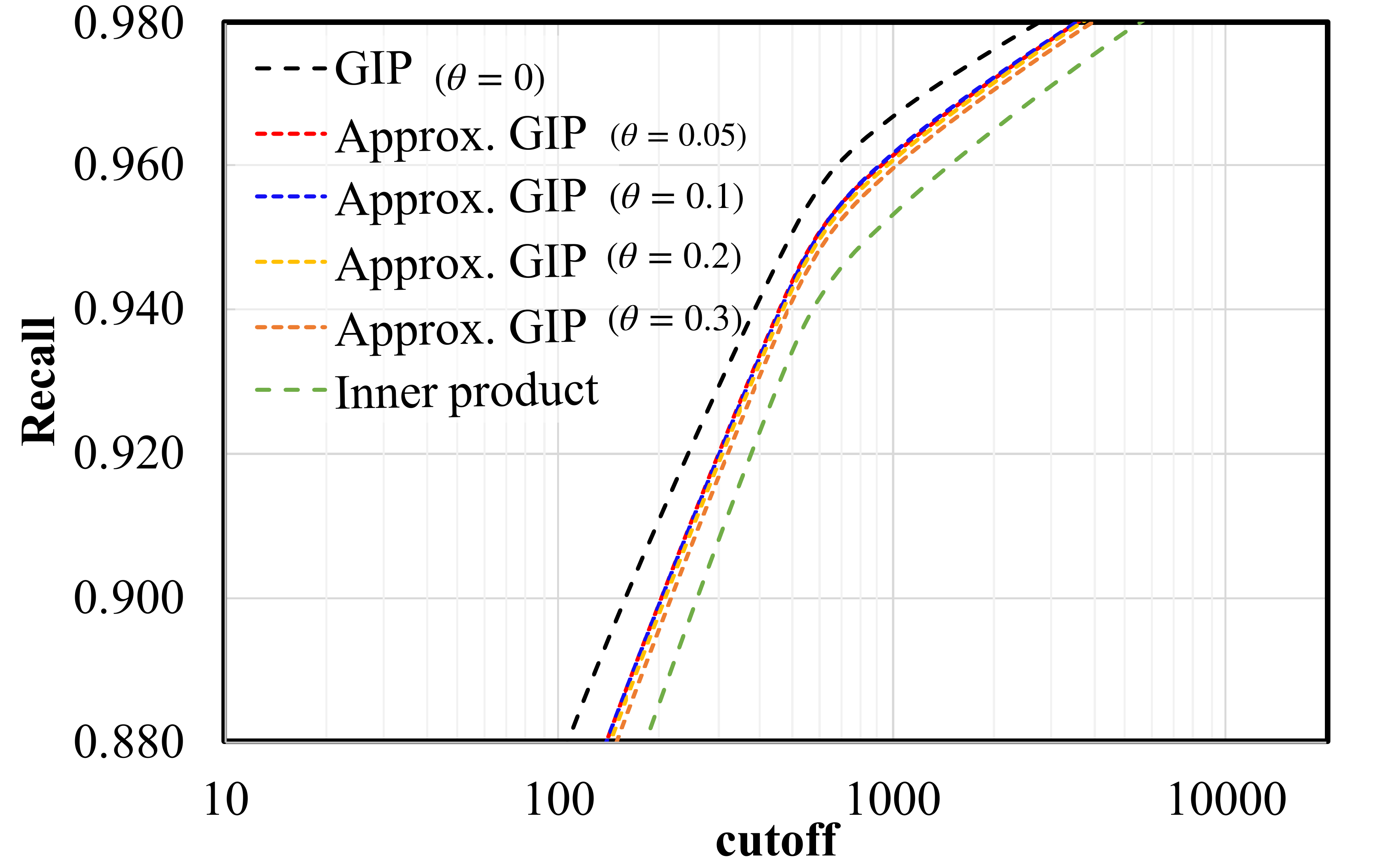}
\end{subfigure}
\caption{Recall comparison of approximate retrieval approaches at different cutoffs. Approx.\ GIP refers to approximate gated inner product with threshold $\theta$. Inner product refers to standard inner product between query and passage value vectors without involving the index vectors. }
\label{fig:retrieval_comparison}
\end{figure}

\begin{figure}[t]
\begin{subfigure}[b]{0.45\columnwidth}
\caption{$\deladedlr(768)$ R@1K}
\includegraphics[width=\columnwidth]{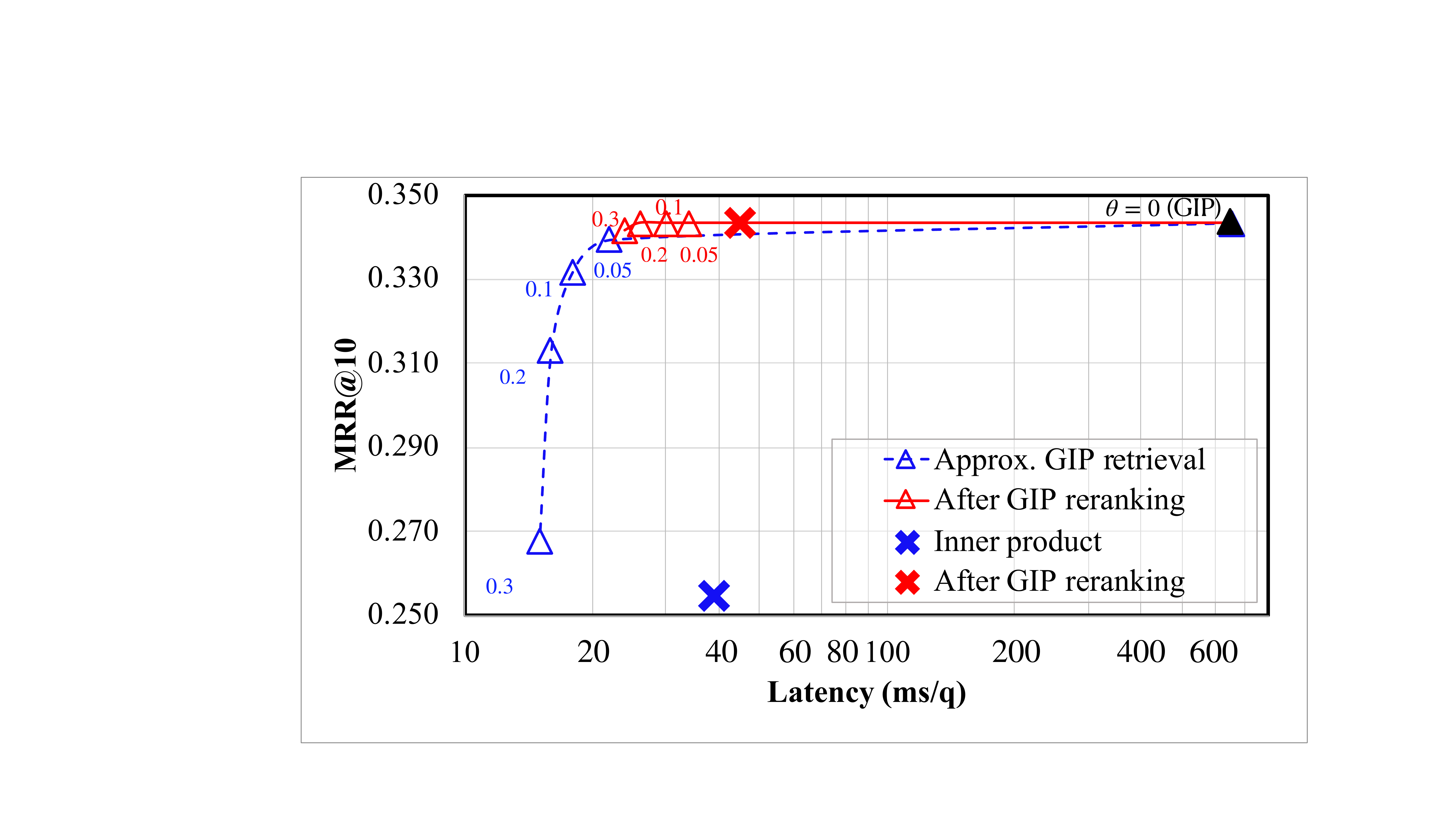}
\end{subfigure}
\begin{subfigure}[b]{0.45\columnwidth}
\caption{$\deladedlr(768)$ MRR@10}
\includegraphics[width=\columnwidth]{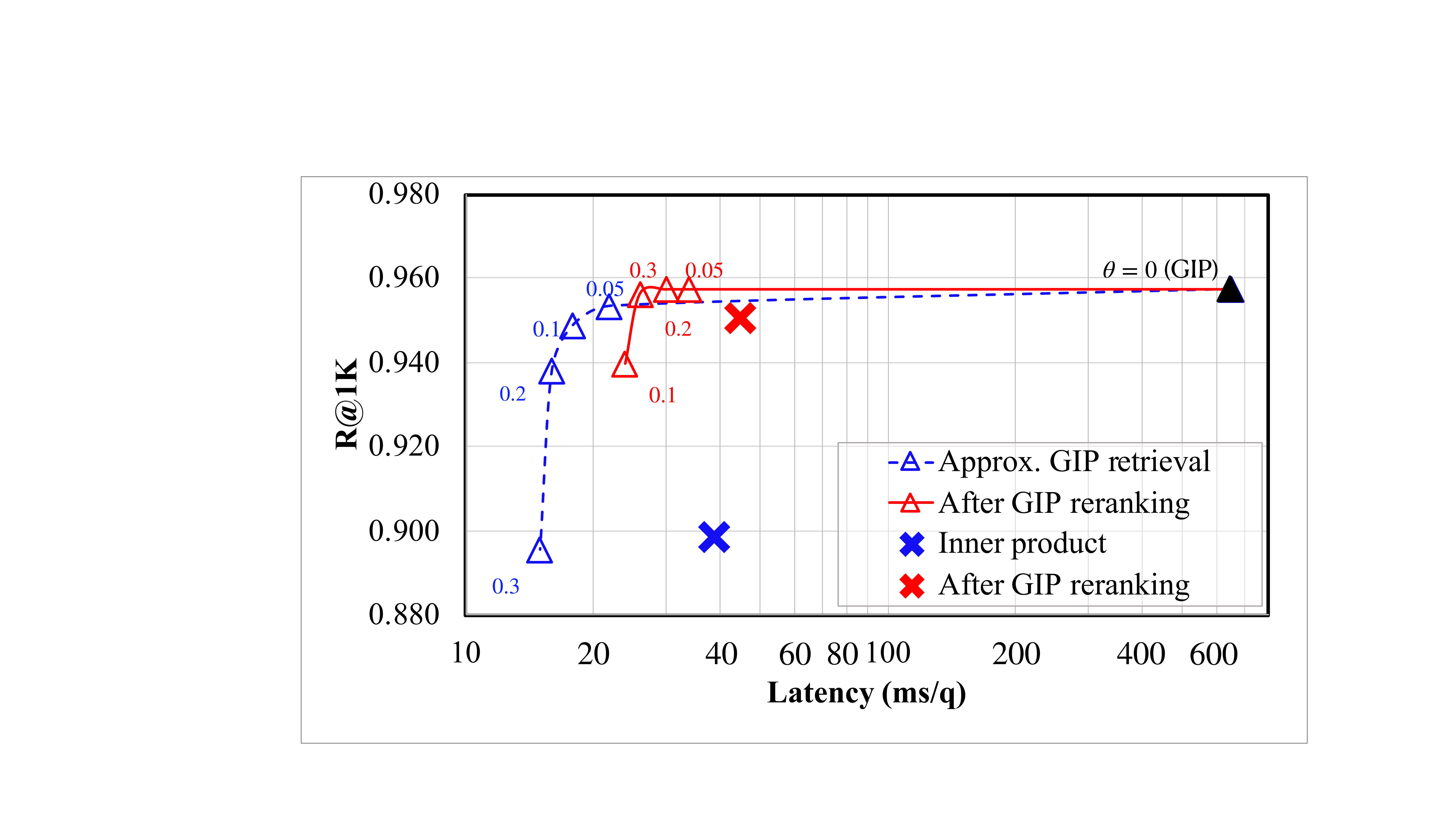}
\end{subfigure}
\begin{subfigure}[b]{0.45\columnwidth}
\caption{$\deladedhr(768)$ R@1K}
\includegraphics[width=\columnwidth]{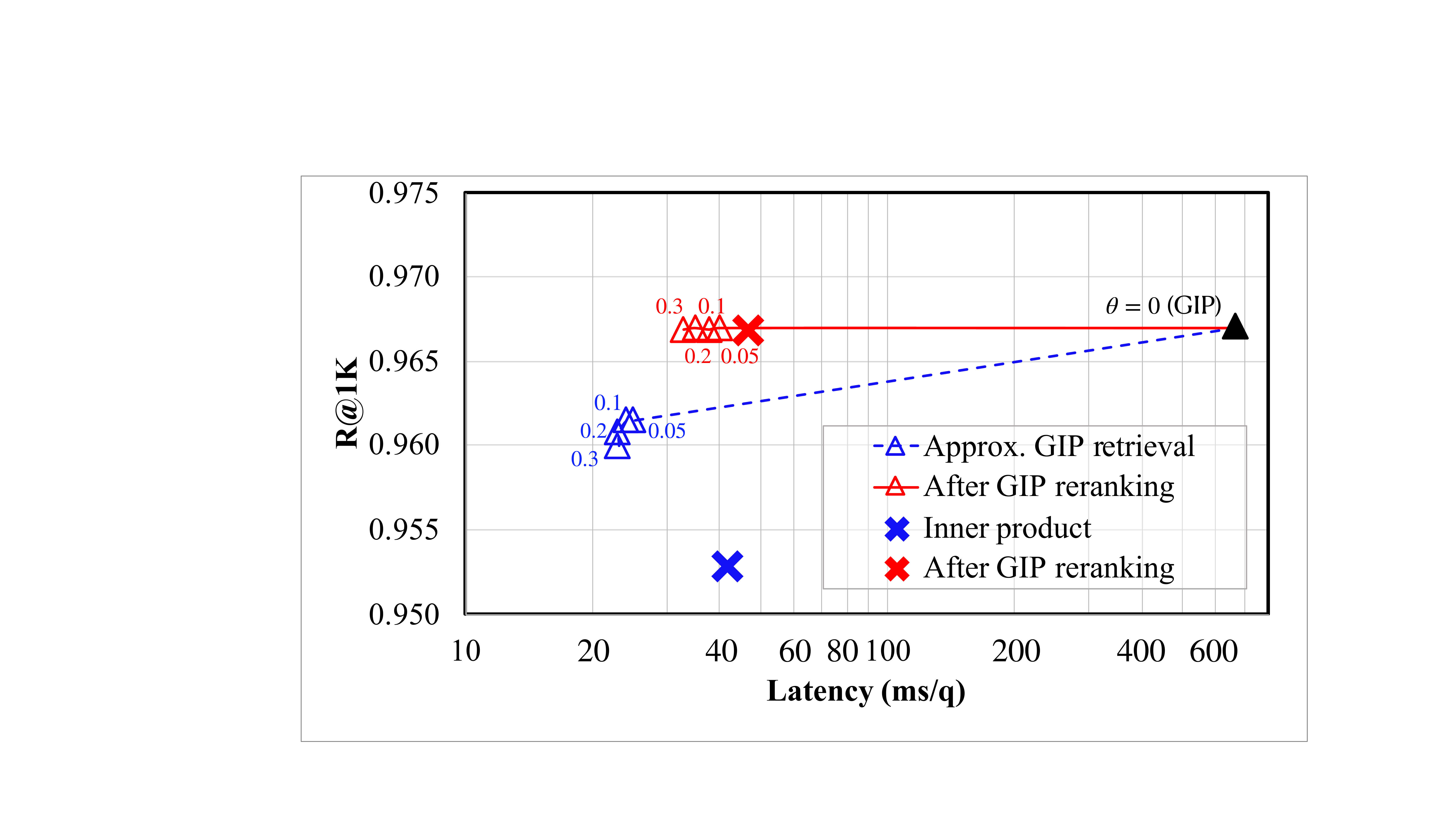}
\end{subfigure}
\begin{subfigure}[b]{0.45\columnwidth}
\caption{$\deladedhr(768)$ MRR@10}
\includegraphics[width=\columnwidth]{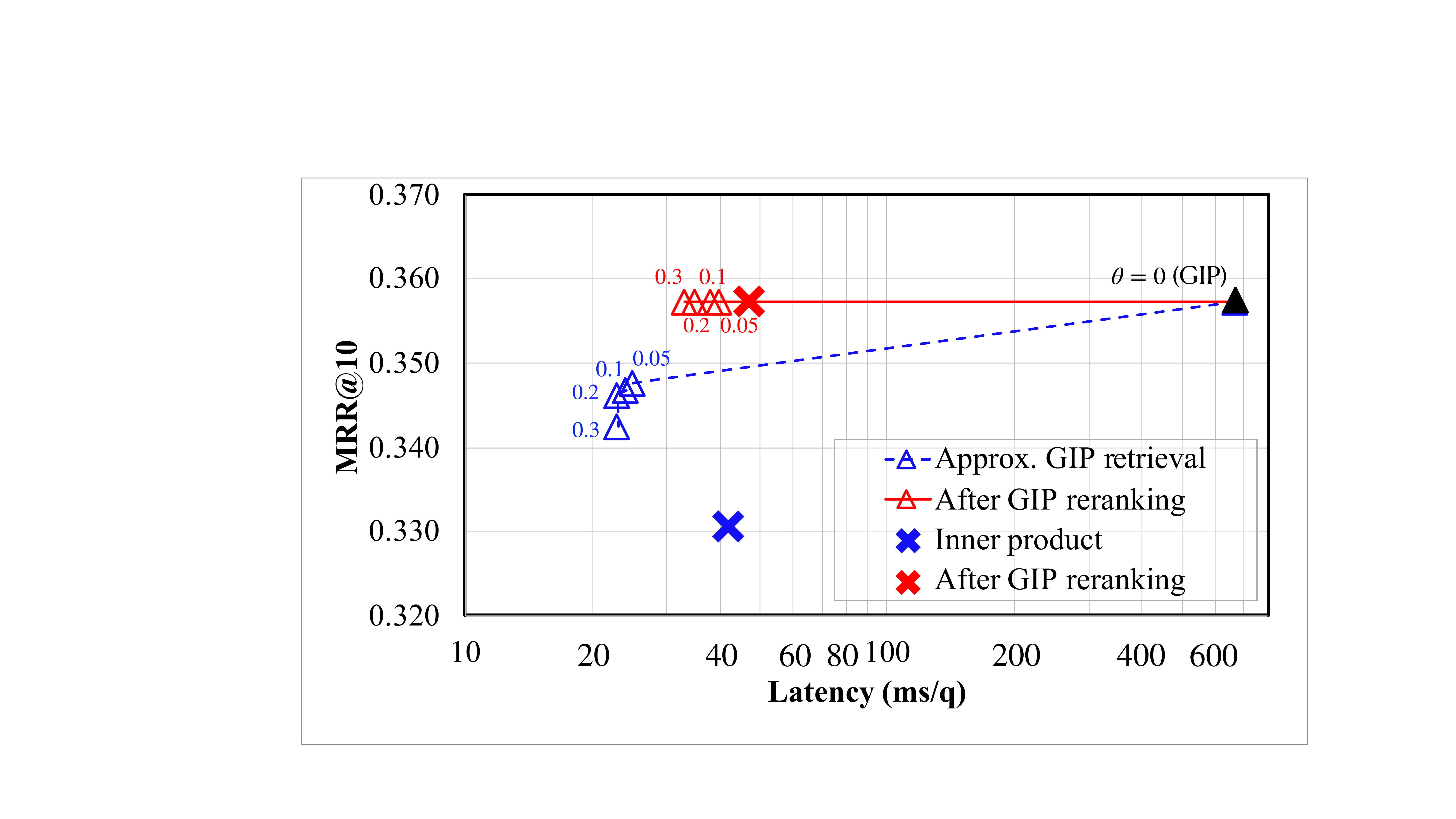}
\end{subfigure}
\caption{Two-stage retrieval performance comparisons. 
We compare retrieval latency and effectiveness between approximate gated inner product (Approx.\ GIP) and standard inner product retrieval (between query and passage value vectors without involving the index vectors) and their performance after GIP reranking the top-10000 retrieved candidates. 
The labeled numbers denote the threshold $\theta$ for Approx. GIP retrieval. 
Note that approx. GIP with $\theta=0$ is equal to GIP retrieval without any approximation.}
\label{fig:two-stage_retrieval_comparison}
\end{figure}

%% file: latency_measurement.tex
\begin{table}[t]
	\caption{End-to-End Retrieval Latency for $\deladedhr(768)$ on MS MARCO (Dev)}
	\label{tb:online_latency_measurement}
	\centering
	\resizebox{0.8\width}{!}{
	\begin{threeparttable}
    \begin{tabular}{lll|cccc|cccc|cc}
	\toprule
 & \multicolumn{2}{c|}{Two-stage retrieval} &\multicolumn{4}{c|}{CPU latency (ms/q)}&\multicolumn{4}{c|}{GPU latency (ms/q)}& \multicolumn{2}{c}{Quality}\\
\cmidrule(lr){2-3}	\cmidrule(lr){4-7} \cmidrule(lr){8-11} \cmidrule(lr){12-13}
& 1st& 2nd& Q Enc.& 1st& 2nd&  Total&Q Enc. &  1st& 2nd&  Total& MRR@10 &R@1K\\
	 \midrule
(1) & GIP& -& 164& 72560& \ \ 0& 72724& 12& 670& \ \ 0& 682& 0.357& 0.967\\
(2) &Approx. GIP& GIP& 164& \ \ 9013& 102& \ \ 9279&12& \ \ 23& 10&  \ \ 45& 0.357& 0.967\\
(3) &Inner Product& GIP& 164& \ \ 3428& 102& \ \ 3694&12&  \ \ 37& 10&  \ \ 59& 0.357& 0.967\\
(4) &Inner Product (w/ \texttt{PQ128})& GIP& 164& \ \ \ \ 562& 102& \ \ \ \ 828& 12& \ \ -\tnote{$\ast$}& 10& \ \ -\tnote{$\ast$}& 0.357& 0.965\\ %541M
(5) &Inner Product (w/ \texttt{PQ64})& GIP& 164& \ \ \ \ 284& 102& \ \ \ \ 650& 12& \ \ -\tnote{$\ast$}& 10& \ \ -\tnote{$\ast$}& 0.357& 0.961\\ %541M
	\bottomrule
	\end{tabular}
		\begin{tablenotes}
	\item[$\ast$] \texttt{Faiss.IndexPQ} does not support GPU search.
    \end{tablenotes}
    \end{threeparttable}
	}

\vspace{0.25cm}

\parbox{\textwidth}{{\small Retrieval latency is measured in terms of ms/q with a single thread and batch size 1.
With the exception of row (1), for each query, we retrieve the top-10000 passages at the first stage and then rerank the candidates using the second stage. 
	Approx.\ GIP refers to approximate gated inner product with threshold $\theta=0.3$. Inner product refers to standard inner product between query and passage value vectors without involving the index vectors.}}

\end{table}